\documentclass[aps,prd,twocolumn,showpacs,superscriptaddress,floatfix,nofootinbib]{revtex4} 

\usepackage{amsmath}
\usepackage{graphicx}
\usepackage{bm}
\usepackage{epsfig}
\usepackage{psfrag}
\usepackage{multirow}
\usepackage{dcolumn}
\usepackage{amssymb}
\usepackage{url}
\usepackage{bold-extra}
\usepackage[shortcuts]{extdash}            
\usepackage{expdlist} 
\usepackage[caption=false,labelfont=bf,font=normalsize]{subfig} 

\newcommand {\plainafb}	{\ensuremath{A_{\mathrm{FB}}}}
\newcommand {\afb}	{{\plainafb}}

\newcommand {\pbar}	{{\ensuremath{\bar p}}}
\newcommand {\ppbar}	{{\ensuremath{p\pbar}}}
\newcommand {\tbar}     {{\ensuremath{\bar t}}}
\newcommand {\ttbar}    {{\ensuremath{t\tbar}}}
\newcommand {\qbar}     {{\ensuremath{\bar q}}}

\newcommand {\pjets}    {{\textrm{+jets}}}
\newcommand {\wdbos}      {{\ensuremath{W\!}-boson}}
\newcommand {\wbprod}      {\wdbos\ production}
\newcommand {\wpj}      {{\ensuremath{W\!+}jets}}
\newcommand {\wpjmath}  {{W\!+\mathrm{jets}}}
\newcommand {\Wcc}      {{\ensuremath{W\!+c\bar{c}}}}
\newcommand {\Wbb}      {{\ensuremath{W\!+b\bar{b}}}}

\newcommand {\lpj}      {{\ensuremath{l\pjets}}}
\newcommand {\epj}      {{\ensuremath{e\pjets}}}
\newcommand {\mpj}      {{\ensuremath{\mu\pjets}}}
\newcommand {\lptj}      {{\ensuremath{l\textrm{+}3\textrm{\,jet}}}}
\newcommand {\lpgefj}      {{\ensuremath{l}+\ensuremath{\geq}4\,jet}}
\newcommand {\getb}      {{\ensuremath{\geq}2 \ensuremath{b} tags}}

\newcommand {\GeV}        {{\ensuremath{\,\textrm {GeV}}}}
\newcommand {\TeV}        {{\ensuremath{\,\textrm {TeV}}}}
\newcommand {\ifb}        {{\ensuremath{\,\textrm {fb}^{-1}}}}

\newcommand {\lep}        {{\ensuremath{\lowercase{l}}}} 
\newcommand {\pt}         {{\ensuremath{p_T}}}
\newcommand {\ptl}         {{\ensuremath{p_T^\lep}}}
\newcommand {\ylep}       {{\ensuremath{y_\lep}}}
\newcommand {\qlep}       {{\ensuremath{q_\lep}}}
\newcommand {\qyl}        {{\ensuremath{\qlep\ylep}}}
\newcommand {\absyl}      {{\ensuremath{\left|\ylep\right|}}}

\newcommand {\ttpt} {{\ensuremath{p_T^\ttbar}}}

\newcommand{\emiss} {{/\!\!\!\!E}}
\newcommand{\met}   {{\ensuremath{\emiss_T}}}
\newcommand{\absmet}   {{\ensuremath{\left|\emiss_T\right|}}}

\newcommand{\lbpt}  {{\ensuremath{p^{\mathrm{LB}}_{T}}}}
\newcommand{\ktmin} {{\ensuremath{k^{\min}_T}}}
\newcommand{\mjj}   {{\ensuremath{M_{jj}}}}
\newcommand{\chisq}  {{\ensuremath{\chi^2}}}
\newcommand{\ptthree}   {{\ensuremath{p_T^\mathrm{3rd}}}}
\newcommand{\mjjmin}   {{\ensuremath{M_{jj}^\mathrm{min}}}}
\newcommand{\disc}   {{\ensuremath{D}}}
\newcommand{\dc}   {{\ensuremath{D_c}}}

\newcommand {\afbl}   {{\ensuremath{\plainafb^\lep}}}

\newcommand {\acr}   {{\ensuremath{A_{\mathrm{CR}}}}}
\newcommand {\dacr}   {{\ensuremath{\Delta\acr}}}
\newcommand {\mttbar} {{\ensuremath{m_\ttbar}}}
\newcommand {\aext}   {{\ensuremath{A_{\mathrm{FB}}^{\lep,\mathrm{ex}}}}}
\newcommand {\atotmc}   {{\ensuremath{A_{\mathrm{FB}}^{\lep,\mathrm{tot}}}}}
\newcommand {\amc}   {{\ensuremath{A_{\mathrm{FB}}^{\lep,\mathrm{pred}}}}}

\newcommand {\spwup}   {\ensuremath{\sigma^+_{\mathrm{CR}}}}
\newcommand {\spwdown}   {\ensuremath{\sigma^-_{\mathrm{CR}}}}

\newcommand {\Ntt}    {{\ensuremath{N_{\ttbar}}}}
\newcommand {\Nw}     {{\ensuremath{N_{\wpjmath}}}}
\newcommand {\Nmj}    {{\ensuremath{N_{\mathrm{MJ}}}}}
\newcommand {\Nother}    {{\ensuremath{N_{\mathrm{OB}}}}} 
\newcommand {\Nsel}    {{\ensuremath{N_{\mathrm{sel}}}}}

\newcommand {\Nfl}    {{\ensuremath{N_F^\lep}}}
\newcommand {\Nbl}    {{\ensuremath{N_B^\lep}}}

\newcommand {\djom}  {{\ensuremath{\Delta \phi (\mathrm{jet}_1,\met)}}}

\newcommand {\DZ}     {{D0}} 

 \newcommand {\pythia}   {{\sc pythia}}
 \newcommand {\alpgen}   {{\sc alpgen}}
  \newcommand {\madgraph}   {{\sc madgraph}}
  \newcommand {\comphep}   {{\sc comphep}}
 \newcommand {\mcatnlo}  {{\sc mc@nlo}}
 \newcommand {\herwig}   {{\sc herwig}}
 
 \newcommand {\resbos}   {{\sc resbos}}
 
 \newcommand {\eg}       {{\rm e.g.}}
 \newcommand {\etal}     {{\it et al.}}
 \newcommand {\this}	{{analysis}} 
 \newcommand {\stat}	{{\ensuremath{\textrm{\thinspace(stat.)}}}}
 \newcommand {\syst}	{{\ensuremath{\textrm{\thinspace(syst.)}}}}
 \newcommand {\nominus} {\phantom{\ensuremath{-}}}
 \newcommand {\noone} {\phantom{\ensuremath{1}}}

\newcommand {\ass} {{at $\sqrt{s}=1.96\TeV$}}

\newcommand {\tablestrut} {\rule{0pt}{2.5ex}}
\newcommand {\smallstrut} {\rule{0pt}{2.3ex}}

\DeclareMathOperator{\sgn}{sgn}

\newcommand{\centercell}[1]{\multicolumn{1}{c}{#1}}
\newcommand{\head}[1]{\centercell{#1}} 
\newcommand{\multihead}[2]{\multicolumn{#1}{c}{#2}}

\newcommand {\riptl} {\ensuremath{\geq20}}
\newcommand {\rlptl} {20--35}
\newcommand {\rmptl} {35--60}
\newcommand {\rhptl} {\ensuremath{\geq60}}
\newcommand {\nulptl} {{\ensuremath{20\leq\ptl<35}}}
\newcommand {\numptl} {{\ensuremath{35\leq\ptl<60}}}
\newcommand {\nuhptl} {{\ensuremath{\ptl\geq60}}}
\newcommand {\lptl} {{\ensuremath{\nulptl\GeV}}}
\newcommand {\mptl} {{\ensuremath{\numptl\GeV}}}
\newcommand {\hptl} {{\ensuremath{\nuhptl\GeV}}}

\newcommand {\accmat}  {\ensuremath{\boldsymbol{A}}}
\newcommand {\migmat}  {\ensuremath{\boldsymbol{M}}}

\begin{document}
\hspace{5.2in} \mbox{FERMILAB-PUB-14-041-E}
\title{Measurement of the forward-backward asymmetry in the distribution of \\
       leptons in $\bm{\ttbar}$ events in the lepton+jets channel}
\date{March 5, 2014}

\affiliation{LAFEX, Centro Brasileiro de Pesquisas F\'{i}sicas, Rio de Janeiro, Brazil}
\affiliation{Universidade do Estado do Rio de Janeiro, Rio de Janeiro, Brazil}
\affiliation{Universidade Federal do ABC, Santo Andr\'e, Brazil}
\affiliation{University of Science and Technology of China, Hefei, People's Republic of China}
\affiliation{Universidad de los Andes, Bogot\'a, Colombia}
\affiliation{Charles University, Faculty of Mathematics and Physics, Center for Particle Physics, Prague, Czech Republic}
\affiliation{Czech Technical University in Prague, Prague, Czech Republic}
\affiliation{Institute of Physics, Academy of Sciences of the Czech Republic, Prague, Czech Republic}
\affiliation{Universidad San Francisco de Quito, Quito, Ecuador}
\affiliation{LPC, Universit\'e Blaise Pascal, CNRS/IN2P3, Clermont, France}
\affiliation{LPSC, Universit\'e Joseph Fourier Grenoble 1, CNRS/IN2P3, Institut National Polytechnique de Grenoble, Grenoble, France}
\affiliation{CPPM, Aix-Marseille Universit\'e, CNRS/IN2P3, Marseille, France}
\affiliation{LAL, Universit\'e Paris-Sud, CNRS/IN2P3, Orsay, France}
\affiliation{LPNHE, Universit\'es Paris VI and VII, CNRS/IN2P3, Paris, France}
\affiliation{CEA, Irfu, SPP, Saclay, France}
\affiliation{IPHC, Universit\'e de Strasbourg, CNRS/IN2P3, Strasbourg, France}
\affiliation{IPNL, Universit\'e Lyon 1, CNRS/IN2P3, Villeurbanne, France and Universit\'e de Lyon, Lyon, France}
\affiliation{III. Physikalisches Institut A, RWTH Aachen University, Aachen, Germany}
\affiliation{Physikalisches Institut, Universit\"at Freiburg, Freiburg, Germany}
\affiliation{II. Physikalisches Institut, Georg-August-Universit\"at G\"ottingen, G\"ottingen, Germany}
\affiliation{Institut f\"ur Physik, Universit\"at Mainz, Mainz, Germany}
\affiliation{Ludwig-Maximilians-Universit\"at M\"unchen, M\"unchen, Germany}
\affiliation{Panjab University, Chandigarh, India}
\affiliation{Delhi University, Delhi, India}
\affiliation{Tata Institute of Fundamental Research, Mumbai, India}
\affiliation{University College Dublin, Dublin, Ireland}
\affiliation{Korea Detector Laboratory, Korea University, Seoul, Korea}
\affiliation{CINVESTAV, Mexico City, Mexico}
\affiliation{Nikhef, Science Park, Amsterdam, the Netherlands}
\affiliation{Radboud University Nijmegen, Nijmegen, the Netherlands}
\affiliation{Joint Institute for Nuclear Research, Dubna, Russia}
\affiliation{Institute for Theoretical and Experimental Physics, Moscow, Russia}
\affiliation{Moscow State University, Moscow, Russia}
\affiliation{Institute for High Energy Physics, Protvino, Russia}
\affiliation{Petersburg Nuclear Physics Institute, St. Petersburg, Russia}
\affiliation{Instituci\'{o} Catalana de Recerca i Estudis Avan\c{c}ats (ICREA) and Institut de F\'{i}sica d'Altes Energies (IFAE), Barcelona, Spain}
\affiliation{Uppsala University, Uppsala, Sweden}
\affiliation{Taras Shevchenko National University of Kyiv, Kiev, Ukraine}
\affiliation{Lancaster University, Lancaster LA1 4YB, United Kingdom}
\affiliation{Imperial College London, London SW7 2AZ, United Kingdom}
\affiliation{The University of Manchester, Manchester M13 9PL, United Kingdom}
\affiliation{University of Arizona, Tucson, Arizona 85721, USA}
\affiliation{University of California Riverside, Riverside, California 92521, USA}
\affiliation{Florida State University, Tallahassee, Florida 32306, USA}
\affiliation{Fermi National Accelerator Laboratory, Batavia, Illinois 60510, USA}
\affiliation{University of Illinois at Chicago, Chicago, Illinois 60607, USA}
\affiliation{Northern Illinois University, DeKalb, Illinois 60115, USA}
\affiliation{Northwestern University, Evanston, Illinois 60208, USA}
\affiliation{Indiana University, Bloomington, Indiana 47405, USA}
\affiliation{Purdue University Calumet, Hammond, Indiana 46323, USA}
\affiliation{University of Notre Dame, Notre Dame, Indiana 46556, USA}
\affiliation{Iowa State University, Ames, Iowa 50011, USA}
\affiliation{University of Kansas, Lawrence, Kansas 66045, USA}
\affiliation{Louisiana Tech University, Ruston, Louisiana 71272, USA}
\affiliation{Northeastern University, Boston, Massachusetts 02115, USA}
\affiliation{University of Michigan, Ann Arbor, Michigan 48109, USA}
\affiliation{Michigan State University, East Lansing, Michigan 48824, USA}
\affiliation{University of Mississippi, University, Mississippi 38677, USA}
\affiliation{University of Nebraska, Lincoln, Nebraska 68588, USA}
\affiliation{Rutgers University, Piscataway, New Jersey 08855, USA}
\affiliation{Princeton University, Princeton, New Jersey 08544, USA}
\affiliation{State University of New York, Buffalo, New York 14260, USA}
\affiliation{University of Rochester, Rochester, New York 14627, USA}
\affiliation{State University of New York, Stony Brook, New York 11794, USA}
\affiliation{Brookhaven National Laboratory, Upton, New York 11973, USA}
\affiliation{Langston University, Langston, Oklahoma 73050, USA}
\affiliation{University of Oklahoma, Norman, Oklahoma 73019, USA}
\affiliation{Oklahoma State University, Stillwater, Oklahoma 74078, USA}
\affiliation{Brown University, Providence, Rhode Island 02912, USA}
\affiliation{University of Texas, Arlington, Texas 76019, USA}
\affiliation{Southern Methodist University, Dallas, Texas 75275, USA}
\affiliation{Rice University, Houston, Texas 77005, USA}
\affiliation{University of Virginia, Charlottesville, Virginia 22904, USA}
\affiliation{University of Washington, Seattle, Washington 98195, USA}
\author{V.M.~Abazov} \affiliation{Joint Institute for Nuclear Research, Dubna, Russia}
\author{B.~Abbott} \affiliation{University of Oklahoma, Norman, Oklahoma 73019, USA}
\author{B.S.~Acharya} \affiliation{Tata Institute of Fundamental Research, Mumbai, India}
\author{M.~Adams} \affiliation{University of Illinois at Chicago, Chicago, Illinois 60607, USA}
\author{T.~Adams} \affiliation{Florida State University, Tallahassee, Florida 32306, USA}
\author{J.P.~Agnew} \affiliation{The University of Manchester, Manchester M13 9PL, United Kingdom}
\author{G.D.~Alexeev} \affiliation{Joint Institute for Nuclear Research, Dubna, Russia}
\author{G.~Alkhazov} \affiliation{Petersburg Nuclear Physics Institute, St. Petersburg, Russia}
\author{A.~Alton$^{a}$} \affiliation{University of Michigan, Ann Arbor, Michigan 48109, USA}
\author{A.~Askew} \affiliation{Florida State University, Tallahassee, Florida 32306, USA}
% ...
%
%
\collaboration{The D0 Collaboration\footnote{with visitors from
$^{a}$Augustana College, Sioux Falls, SD, USA,
$^{b}$The University of Liverpool, Liverpool, UK,
$^{c}$DESY, Hamburg, Germany,
$^{d}$Universidad Michoacana de San Nicolas de Hidalgo, Morelia, Mexico
$^{e}$SLAC, Menlo Park, CA, USA,
$^{f}$University College London, London, UK,
$^{g}$Centro de Investigacion en Computacion - IPN, Mexico City, Mexico,
$^{h}$Universidade Estadual Paulista, S\~ao Paulo, Brazil,
$^{i}$Karlsruher Institut f\"ur Technologie (KIT) - Steinbuch Centre for Computing (SCC),
D-76128 Karlsrue, Germany,
$^{j}$Office of Science, U.S. Department of Energy, Washington, D.C. 20585, USA,
$^{k}$American Association for the Advancement of Science, Washington, D.C. 20005, USA
$^{l}$Kiev Institute for Nuclear Research, Kiev, Ukraine
and
$^{m}$Laboratoire de Physique Theorique, Orsay, FR.
}} \noaffiliation
\vskip 0.25cm

%
%
%
\begin{abstract}
We present measurements of the forward-backward asymmetry in the angular distribution of leptons
from decays of top quarks and antiquarks produced in proton-antiproton collisions. 
We consider the final state containing a lepton and at least three jets.
The entire sample of data collected by the \DZ\ experiment
during Run II of the Fermilab Tevatron Collider, corresponding to $9.7\ifb$ of integrated luminosity, is used.
The asymmetry measured for reconstructed leptons 
is $\afbl = \big(2.9 \pm 2.1 \stat ^{+1.5}_{-1.7} \syst \big)$\%. 
When corrected for efficiency and resolution effects within the lepton rapidity coverage of $\absyl<1.5$,
the asymmetry is found to be $\afbl = \big(4.2 \pm 2.3 \stat ^{+1.7}_{-2.0} \syst \big)$\%.
Combination with the asymmetry measured in the dilepton final state yields
$\afbl = \big(4.2 \pm 2.0 \stat \pm 1.4 \syst \big)$\%.
We examine the dependence of \afbl\ on the transverse momentum and rapidity of the
lepton.
The results are in agreement with predictions from the next-to-leading-order 
QCD generator \mcatnlo, which predicts an asymmetry of $\afbl = 2.0$\% for $\absyl<1.5$.
\end{abstract}

%
%

\maketitle

\section{Introduction}
Within the standard model of particle physics (SM), top quarks are usually
produced via quantum chromodynamic (QCD) interactions in
quark-antiquark pairs.
The process $\ppbar\to\ttbar(X)$ is predicted to produce more events for which the rapidity
of the top quark, $y_t$, is greater than the rapidity of the top antiquark,
$y_\tbar$, than events for which $y_t$ is less than $y_\tbar$.
The rapidity $y$ is defined as 
$y\left(\theta,\beta\right)=\frac{1}{2}\ln\left[\left(1+\beta\cos\theta\right)/\left(1-\beta\cos\theta\right)\right]$,
where $\theta$ is the polar angle and $\beta$ is the ratio of a particle's momentum to its energy.
The angle $\theta=0$ corresponds to the direction of the incoming proton.
This predicted forward-backward asymmetry is mostly due to
contributions at order $\alpha_s^3$, where $\alpha_s$ is the QCD coupling constant~\cite{bib:k_and_r}.
There are also smaller contributions to the forward-backward asymmetry
from electroweak (EW) interactions~\cite{bib:SMEW}.
Forward-backward asymmetries in $\ppbar\to\ttbar(X)$ production previously measured at 
the Fermilab Tevatron Collider~\cite{bib:ourPRD,bib:D0dilep,bib:CDFdep}
were found to be somewhat higher than the SM predictions~\cite{bib:Bern,bib:predAFB}.
With a mass of approximately $173\GeV$~\cite{bib:mtop}, the top quark is the most massive known elementary particle,
which raises the possibility that the asymmetry is enhanced by effects beyond the SM.
Hence, the previously measured asymmetries
led to studies of possible causes not only within, but also beyond the SM~\cite{bib:review}.

Top quarks decay almost exclusively into a $b$ quark and a $W$ boson, and $W$ bosons decay
either hadronically to a quark and an antiquark or leptonically to a lepton and a neutrino. 
Thus, \ttbar\ events are usually classified based on the number of leptons from 
the decays of the $W$ bosons into the dilepton, lepton+jets (\lpj), and all-jets channels.
The \ttbar\ production asymmetry was first measured by the \DZ\ Collaboration~\cite{bib:p17PRL} in the \lpj\ channel.
The result of Ref.~\cite{bib:p17PRL} was superseded by that of Ref.~\cite{bib:ourPRD}, 
where a dataset corresponding to an integrated luminosity of 5.4\ifb\
was used to measure an inclusive asymmetry of $\left(20^{+6}_{-7}\right)\%$.
The CDF Collaboration measured this asymmetry in the
\lpj\ channel with $9.4\ifb$ of integrated luminosity, 
finding $\afb=\left(16.4\pm4.5\right)\%$~\cite{bib:CDFdep}.
These measured values can be compared to SM predictions, 
for example, $\afb=\left(8.8\pm0.9\right)\%$~\cite{bib:Bern}.

The forward-backward asymmetry in the production of \ttbar\ pairs leads to a forward-backward 
asymmetry \afbl\ in the angular distribution of the leptons produced in the \ttbar\ decays~\cite{bib:observable}.
The asymmetry \afbl\ was first measured by \DZ~\cite{bib:ourPRD} as a
cross-check of the asymmetry of the \ttbar\ pair
and to demonstrate that the observed tension with the SM
should not be attributed to biases introduced by the algorithm used to reconstruct the \ttbar\ system 
or to the corrections for the detector acceptance and resolution effects. 
In SM $\ppbar\to\ttbar(X)$ production, the polarization of top quarks
is negligible and the leptons are produced isotropically
(in the appropriate reference frames) leading to an \afbl\ that is smaller than \afb.
The \DZ\ Collaboration measured $\afbl=(15.2 \pm 4.0)\%$ in the \lpj\ channel for $\absyl<1.5$, where \ylep\ is
the rapidity of the lepton from top quark decay~\cite{bib:ourPRD} and
$\afbl=(5.8 \pm 5.3)\%$ in the dilepton channel for $\absyl<2$~\cite{bib:D0dil}.
The CDF Collaboration measured this asymmetry in the \lpj\ channel
and found values extrapolated to the full acceptance of $\afbl=(9.4^{+3.2}_{-2.9})\%$~\cite{bib:CDFllj}. 
The corresponding SM predictions range from $2.0\%$ to $3.8\%$~\cite{bib:mcatnlo,bib:MCFMttbar,bib:Bern}.
The higher predictions include electroweak corrections, which increase
\afbl\ by less than a percent (absolute).
The dominant uncertainty on these predictions is from the renormalization and factorization scales,
and is evaluated to be up to $1.0\%$~\cite{bib:MCFMttbar,bib:Bern}.
The results of the previous measurements could be taken as an indication of effects beyond the SM that 
lead to the production of polarized top quarks~\cite{bib:top_pol}.
Motivated by the desire to further investigate this tension and by the potential sensitivity of
\afbl\ to new physics, we pursue this analysis in greater detail and with a larger dataset.

Measuring the leptonic asymmetry rather than the \ttbar\ asymmetry
has additional benefits.
The measurements of the \ttbar\ asymmetry require full reconstruction of the \ttbar\ decay
chain, accomplished by assuming on-shell top quarks that each decay to three final state fermions.
These assumptions limit the validity of a comparison of data to calculations that include 
higher orders in top quark decay and off-shell top quarks (\eg, in loops).
These limitations are not intrinsic to the lepton-based asymmetry. Although we make some use
of \ttbar\ reconstruction, the effects of off-shell top quarks and of decays with
additional final-state partons on the measurement of \afbl\ are negligible.

Experimentally, 
the direction of a lepton is determined with far greater precision than that of a top quark.
Thus, corrections for the detector acceptance and experimental resolutions are simpler.
Furthermore, with no need for full reconstruction of the \ttbar\ system, 
the \lptj\ sample can be used for this measurement in addition to the previously used \lpgefj s sample. 
This addition almost doubles the number of \ttbar\ events analyzed, at
the expense of a lower signal-to-background ratio. 
The inclusion of the \lptj\ sample also reduces the acceptance corrections, 
which are a leading source of systematic uncertainty in Ref.~\cite{bib:ourPRD}.

In Ref.~\cite{bib:CDFdep}, the CDF Collaboration reported a strong increase of \afb\ with the 
invariant mass of the \ttbar\ system, \mttbar. 
The dependence of the asymmetry on \mttbar\ observed in the previous \DZ\ measurement~\cite{bib:ourPRD}
is statistically consistent with both the SM prediction and the CDF result. 
Measuring \mttbar\ requires full reconstruction of the \ttbar\ system,
but we can also study the dependence of the asymmetry on the \ttbar\ kinematics
by relying on the transverse momentum of the lepton, \ptl.
This observable can readily be studied in \lptj\ events and is
measured with far greater precision than \mttbar.
Furthermore, \ptl\ is strongly correlated with \mttbar, and is useful
in comparing data to the predictions of different models~\cite{bib:theory_leptons}.
This differential measurement is therefore well motivated both
experimentally and as a test of new physics models.

We report here an updated measurement of \afbl, using the full dataset collected
by the \DZ\ experiment during Run II of the Fermilab Tevatron Collider \ass.
We extend the measurement to include \lptj\ events, which required
improvements in the background modeling, and measure the \ptl\ dependence of
\afbl\ for the first time.
The measurement reported in this paper supersedes the results of Ref.~\cite{bib:ourPRD}.
%
%
\section{The \DZ\ detector}
The \DZ\ detector~\cite{bib:d0det} is a multipurpose particle detector
with the following main components.
At the core is a silicon microstrip tracker~\cite{bib:SMT}, surrounded
by a scintillating-fiber tracker.
The tracking detectors are located within a superconducting solenoid magnet that
provides a $2\,$T axial magnetic field and is surrounded by liquid argon / uranium calorimeters. 
The calorimeters~\cite{bib:d0calorimetry} are enclosed within a central barrel cryostat and two end-cap cryostats.
A muon detection system~\cite{bib:d0muons} surrounds the calorimetry, and consists of an iron toroidal magnet, 
tracking detectors, and scintillation trigger detectors.
Data collection is triggered by a three-level system.

The \DZ\ coordinate system has the $z$ axis along the direction of the
proton beam. Particle directions are presented in terms of their
azimuthal angle $\phi$ and their
pseudorapidity $\eta=-\ln\left[\tan\left(\frac{\theta}{2}\right)\right]$,
where $\theta$ is the polar angle.

%
%
\section{Defining the lepton-based asymmetry}

In this \this, we measure the charge (\qlep) and rapidity (\ylep) of 
the electron or muon that originates from the $W$ boson from top quark decay.
Events with $\qyl>0$ are defined as forward and events with $\qyl<0$
are defined as backward. We define the lepton-based forward-backward asymmetry as
\begin{equation}
\afbl = \frac {\Nfl - \Nbl} {\Nfl + \Nbl},
\label{eq:afbl}
\end{equation}
where $\Nfl$ and $\Nbl$ are the number of forward and backward events, respectively.
All asymmetries are reported after subtracting the estimated background.

The asymmetry can be defined at the ``reconstruction level,'' which refers
to the measured lepton parameters and is affected by acceptance and resolution.  
To enable direct comparisons with SM and non-SM calculations, 
the asymmetry can also be defined at the ``production level,'' before acceptance
and resolution effects take place. 
The production level is sometimes also denoted as the generator level, or the parton level. 

Though the rapidity coverage differs for electrons and muons, we assume
lepton flavor universality and define \afbl\ and the acceptance in terms of \absyl.
To avoid large acceptance corrections, only events with $\absyl<1.5$ are used
(see Ref.~\cite{bib:ourPRD}). 
Throughout most of this paper, the production\-/level \afbl\ is defined counting
only leptons produced within this lepton coverage.
However, in Section~\ref{sec:comb} 
we also discuss asymmetries extrapolated to the full acceptance.

%
%
\section{Analysis strategy}
\label{sec:strategy} 
The selection focuses on $\ttbar(X)\to W^{+}bW^{-}\bar{b} (X)$ events in the
\lpj\ decay mode,  where one $W$ boson decays hadronically ($\qbar q'$)
and the other decays leptonically ($l\bar{\nu_{l}}$). 
The experimental signature of this decay mode is one isolated lepton ($e$ or $\mu$) with 
a large \ptl, a significant imbalance in transverse momentum (\met, with the letter $E$ indicating 
a calorimetry-based observable) from 
the undetected neutrino, and jets arising from the two $b$ quarks and
from the two quarks from $W\to\qbar q'$ decay.
We select electrons and muons, which arise either directly from the $W$ boson
decay or through an intermediate $\tau$ lepton. 

A prototypical \lpj\ event contains four final state quarks and
hence four jets. Previous measurements selected events with at least four jets. 
Only half of the \ttbar\ events in the \lpj\ channel have four or more
selected jets, as one of the jets may fail the selection criteria 
due to insufficient transverse momentum (\pt) or due to large absolute rapidity.
In addition, the decay products of two of the final state partons may be clustered into a single jet.

In this measurement we also select events with three jets.
The inclusion of three-jet events has the advantages of increasing the 
statistical power of the measurement and making the measurement less susceptible to biases from selection.  
However, these additional events have a lower signal-to-background
ratio than the events with $\ge 4$ jets.
To maximize the statistical power of the purer subsets, we separate the measurement into 
several channels, defined by the number of jets (3 or $\ge 4$) and the
number of ``$b$-tagged'' jets (0, 1, or $\ge 2$), that is, jets
identified as likely to originate from a $b$ quark.
 
We identify variables that discriminate between the \ttbar\ signal and
the production of $W$ bosons in association with jets (\wpj),
and combine the variables into a single discriminant \disc\ by neglecting the
small correlations between them.
There are separate discriminants for \lptj\ events and \lpgefj\ events.
We use these discriminants to estimate the number of selected \ttbar\ events and their 
reconstruction\-/level \afbl\ (see Section~\ref{sec:reco_afb}).
 
The addition of three jet events increases the sensitivity of the analysis to the 
modeling of \wpj\ production, which contributes most of the selected three jet events
but only a minority of the selected $\ge$4 jet events.
We study and improve the modeling of \afbl\ in the \wpj\ background 
using a top-depleted control sample (see Section~\ref{sec:wcalib}).

We then correct the \qyl\ distribution to the production level within the lepton coverage,
and measure the production\-/level \afbl. 
Since the angular resolutions for electrons and muons are excellent,
the incorrect classification of events as forward or backward is negligibly small.
We therefore correct \afbl\ only for acceptance effects (see Section~\ref{sec:unfold}). 

In addition to measuring the inclusive \afbl, we also measure \afbl\ in three \ptl\ regions: 
\lptl, \mptl, and \hptl. 
To measure the \ptl\ dependence, we first correct for migrations between different \ptl\ 
regions and then correct for the effects of acceptance (see Section~\ref{sec:unfold}).

%
%
\section{Event selection}
\label{sec:selection}
The event selection criteria used in this \this\ are similar to those used to measure 
the \ttbar\ production cross section in the \lpj\ channel~\cite{bib:D0xsect}.
In particular, we also accept events with three selected jets.
The reconstruction and identification of jets, isolated leptons, and \met\ is described
in Ref.~\cite{bib:objid}.

Only jets with transverse momentum $\pt>20\GeV$ and $|\eta|<2.5$ are considered for further analysis, 
and events are required to contain at least three such jets.
The leading jet, that is the jet with the largest \pt, is also required to have $\pt>40\GeV$. 
As in Ref.~\cite{bib:D0xsect}, we minimize the effect of multiple \ppbar\ collisions in the same
bunch crossing by requiring that jets are vertex confirmed~\cite{bib:D0vc}, i.e., have at least two tracks
within the jet cone pointing back to the primary \ppbar\ collision vertex (PV).

A typical decay of a $b$ hadron occurs at a distance of the order of
1\,mm from the PV and results in charged tracks that are detected by
the tracking system and form a displaced secondary vertex.
Thus, jets that originate from a $b$ quark can be identified by the properties of the
tracks reconstructed within the jet cone, in particular by their displacement from the PV, 
and by the reconstruction of displaced secondary vertices.
Several observables useful for identifying such jets
are combined into a multivariate discriminant~\cite{bib:btagging}
that is used in this analysis to tag $b$ jets by selecting jets likely to originate from a $b$ quark
among the three or four jets with the highest transverse momentum.

The \epj\ and \mpj\ channels have similar event selection requirements.  
Only events collected with single-lepton or lepton+jet triggers are used.
The criteria for selecting \epj\ events are: 
\begin{itemize}
\item one isolated electron with $\pt>20\GeV$ and $|\eta| < 1.1$,
\item $\absmet>20\GeV$, and
\item $\Delta\phi(e,\met) > \left(2.2 - 0.045 \cdot \met / \mathrm{GeV} \right)$ radians.
\end{itemize}
For \mpj\ events, the criteria are:
\begin{itemize}
\item one isolated muon with $\pt>20\GeV$ and $|\eta| < 1.5$,
\item $25\GeV < \absmet<250\GeV$, and
\item $\Delta\phi(\mu,\met) > \left(2.1 - 0.035 \cdot \met  / \mathrm{GeV} \right)$ radians.
\end{itemize}
Events with a second isolated electron or muon passing the selection are rejected. 
The $\absmet<250\GeV$ cut suppresses events where the
\met\ is due to a mismeasurement of the \pt\ of the muon, which is
reconstructed from the curvature of the track of the muon.
The $\Delta\phi(l,\met)$ cuts reduces the background from QCD multijet (MJ) production.
MJ events can pass the selection when a jet is misidentified as an isolated lepton.
This often results in spurious reconstructed \met\ along the lepton's direction.

In addition to the above criteria of Ref.~\cite{bib:D0xsect}, we also require that the curvature 
of the track associated with the lepton is well measured. 
This requirement, while $\approx97\%$ efficient for leptons produced in \ttbar\ decay,
suffices to lower the lepton charge misidentification rate to less than one part in a thousand.
It also reduces the migration of events among the three \ptl\ regions.

For events with muons with $\pt > 60\GeV$, we also require that the magnitude of the
vector sum of the muon momentum and missing transverse energy is greater than $20\GeV$. 
This requirement rejects events consistent
with low energy muons from low energy jets that are badly reconstructed as having high \pt, 
leading to their misclassification as isolated leptons. 
Such events are part of the MJ background, but their
modeling as part of that background, using the technique described in Section~\ref{sec:fit},
is problematic. To limit any possible mismodeling, we also suppress these events with
additional requirements on the track associated with the muon.
Leptons from signal events pass these additional requirements with $\approx85\%$ efficiency.

The main background after this event selection is due to \wpj\ production.  
There is a smaller contribution from MJ production.
Other small backgrounds from single top quark, $Z$+jets and diboson production are also present.

We use the \mcatnlo\ event generator~\cite{bib:mcatnlo} combined with \herwig\ showering~\cite{bib:herwig} 
to model the behavior of \ttbar\ events, and
\alpgen~\cite{bib:alpgen} combined with \pythia~\cite{bib:pythia} to simulate the \wpj\ background. 
The rate of inclusive \Wcc\ and \Wbb\ production predicted by \alpgen\ 
is scaled up by a factor of $1.47$, so that the ratio of the heavy flavor production rate to the 
inclusive \wpj\ production rate agrees with the ratio calculated at next-to-leading-order 
(NLO)~\cite{bib:D0xsect,bib:mcfm}.
The simulated \wdbos\ \pt\ distribution is reweighted to match the product of the measured $Z$-boson 
\pt\ distribution from \DZ\ data~\cite{bib:Zmumu} and the SM ratio of the distributions of the \wdbos\ \pt\ 
and the $Z$-boson \pt, as calculated at NLO with \resbos~\cite{bib:resbos}.
For the other backgrounds, $Z$+jets events are simulated with \alpgen, 
diboson events are simulated with \pythia\ and events from single top quark production
are simulated with \comphep~\cite{bib:comphep}. The normalizations for the last three background
processes are taken from NLO calculations~\cite{bib:mcfm}.
For all simulated events, event generation is 
followed by the \DZ\ detector simulation and reconstruction programs. 
To model energy depositions from noise and additional \ppbar\ collisions within the same bunch crossing, 
simulated events are overlaid with data from random \ppbar\ crossings.
The properties of the MJ background are evaluated using control samples from collider data. 
%
%
\section{The predicted asymmetries}
\label{sec:preds}

As the asymmetry first appears at order $\alpha_s^3$, with the 
largest contribution due to a loop diagram,
it is not fully simulated by tree\-/level event generators.
In addition, the modeling of selection and reconstruction effects requires full Monte Carlo (MC) simulation.
The \mcatnlo\ event generator is well suited for this measurement as it couples 
a NLO calculation of \ttbar\ production with 
subsequent parton showers to fully simulate \ttbar\ events.
The asymmetries of simulated \ttbar\ and \wpj\ production
are listed in Table~\ref{tab:preds} for different jet and $b$-tag multiplicities.

The two leading contributions to the \ttbar\ asymmetry, at order $\alpha_s^3$, are as follows~\cite{bib:k_and_r}.
In events with no additional radiated partons, 
the interference between the Born and box diagrams leads to a positive asymmetry.
At order $\alpha_s^3$, the \pt\ of the \ttbar\ system for these events is $\ttpt=0$.
However, realistic simulation of these processes using parton showers allows for some transverse boost
of the \ttbar\ system.
The interference between diagrams containing initial or final state radiation decreases the asymmetry.
These events have non-zero \ttpt\ at order $\alpha_s^3$, 
and \ttpt\ is usually increased by the parton showers.
Thus the predicted asymmetry decreases as a function of the \pt\ of the \ttbar\ system, \ttpt~\cite{bib:ourPRD}, 
which can also be seen here as a decrease with increasing jet multiplicity. 

In the case of \wpj\ background production, 
$W$ bosons produced by interactions involving gluons or sea quarks contribute positively to the asymmetry. 
On the other hand, 
$W$ bosons produced by valence-valence collisions contribute negatively to the overall asymmetry. 
The production of $W$ bosons in
association with heavy flavor quarks occurs predominantly due to valence--valence
collisions, and thus has a lower \afbl\ compared to inclusive \wbprod, 
as seen in Table~\ref{tab:preds}.

\begin{table}[htbp]
\caption{Simulated reconstructed asymmetries for selected \ttbar\ and \wpj\ events, by event category.
The quoted uncertainties are due to the finite sizes of the simulated samples.  \label{tab:preds}
}
\begin{ruledtabular} 
\begin{tabular}{lcc}
                &  \multihead{2}{\hspace{-5ex}\afbl, \%}\\
 \head{Channel} &  \head{\ttbar\ signal} & \head{\wpj\ background}\\
\hline
\lpj, all channels & \nominus$1.6\pm0.1$ &$13.0 \pm0.2$ \\
\hline
\lptj s, 0 $b$ tags   & \nominus$2.3\pm0.3$ &      $13.5 \pm0.3$ \\
\lptj s, 1 $b$ tag    & \nominus$2.7\pm0.3$ &      $11.6 \pm0.4$ \\
\lptj s, \getb        & \nominus$2.8\pm0.2$ & \noone$7.4 \pm0.9$ \\
\hline
\lpgefj s, 0 $b$ tags &        $-0.9\pm0.4$ &      $14.1 \pm0.9$ \\
\lpgefj s, 1 $b$ tag  & \nominus$0.5\pm0.2$ &      $14.5 \pm1.0$ \\
\lpgefj s, \getb      & \nominus$1.1\pm0.2$ & \noone$8.8 \pm1.9$ \\
\end{tabular}
\end{ruledtabular}
\end{table}

%
%
\section{\boldmath Measuring the reconstructed \afbl}
\label{sec:reco_afb}
We construct a discriminant (see Section~\ref{sec:disc}), and extract the sample composition and
the asymmetry using a maximum likelihood fit to the distribution of
the discriminant and the distribution of the sign of \qyl\ (see Section~\ref{sec:fit}).
The asymmetry values measured at this stage rely on the 
simulated asymmetry of the \wpj\ background.
We then use the estimated sample composition to derive weights for the
simulated \wpj\ background that are based on the asymmetry of control data, as described in Section~\ref{sec:wcalib}.
We apply this reweighting, which does not affect the estimation of the sample
composition, and repeat the maximum likelihood fit to measure the reconstructed \afbl\ for \ttbar\ events.
%
%
\subsection{The discriminant}
\label{sec:disc}

We choose input variables that 
(a) provide good separation between \ttbar\ signal and \wpj\ production, (b) are well modeled,
and (c) have little correlation with each other, and with \ylep, \qlep, and \ptl.
We combine the input variables to form a discriminant \disc\ (as in Refs.~\cite{bib:p17PRL,bib:Whel})
based on the approximate likelihood ratio between the \ttbar\ and \wpj\ hypotheses.
Small correlations between the input variables are neglected,
so only their simulated one-dimensional distributions enter \disc.

For the \lpgefj\ channels, \disc\ is constructed exactly as in Ref~\cite{bib:ourPRD}.
We first reconstruct the full \ttbar\ decay chain using a constrained kinematic fit algorithm~\cite{bib:hitfit}.
For each assignment of the four leading jets to the four quarks from \ttbar\ decay,
the algorithm scales the four-momenta of the observed objects to minimize a \chisq\ test statistic. 
The \chisq\ test statistic measures the consistency of the scaled four-momenta
with the constraints imposed by the known $W$ boson and top quark masses, given the experimental resolutions.
Only assignments that are consistent with the observed $b$ tags are considered.
The most likely assignment and the scaled four-momenta that minimize \chisq\
are used to reconstruct the \wdbos\ and top-quark resonances.
We then build the discriminant from the following variables:
\begin{itemize}
\item \chisq\ of the likeliest assignment.
Low values indicate a \ttbar\ event.
\item \lbpt, the transverse momentum of the leading $b$-tagged jet, 
or when no jets are $b$ tagged, the \pt\ of the leading jet.
Values below $\lbpt\approx50\GeV$ are indicative of \wpj\ production.
\item $\ktmin=\min\left(p_{T,a},p_{T,b}\right)\cdot\Delta{\cal R}_{ab}$, where
$\Delta{\cal R}_{ab}=\sqrt{\left(\eta_a-\eta_b\right)^2+\left(\phi_a-\phi_b\right)^2}$ 
is the angular distance between the two closest jets, $a$ and $b$, and $p_{T,a}$ and $p_{T,b}$ are 
their transverse momenta.
\item \mjj, the invariant mass of the jets assigned to the $W\to q \qbar'$ decay 
in the kinematic fit, calculated using kinematic quantities before the fit.
\end{itemize}
Of these variables, only \chisq\ depends on the lepton, and that
dependence is small as it also depends on the kinematics of the four leading jets.
Thus, this discriminant has little correlation with \ptl\ and \qyl.

The variables \chisq\ and \mjj\ are based on the full \ttbar\ reconstruction, 
so for the \lptj\ channels we construct a different discriminant.
It is constructed in the same manner, but with the following variables:
\begin{itemize}
\item $S$, the sphericity, defined as 
  $S = \frac{3}{2} (\lambda_2 + \lambda_3)$, where  $\lambda_2$ and $\lambda_3$ are the largest two out of 
  the three eigenvalues of the normalized quadratic momentum tensor $M$. The tensor $M$ is defined as
  \begin{equation}
    M_{ij} = \frac {\sum_o {p_i^o p_j^o}}  {\sum_o{|p^o|^2 }},
    \label{eq:tensor}
  \end{equation}
  where $p^o$ is the momentum vector of a reconstructed object $o$, and $i$ and $j$ are the 
  three Cartesian coordinates. 
  The sum over objects includes the three selected jets and the selected charged lepton. 
  Due to the high mass of the top quarks, \ttbar\ events tend to be more spherical than background events. 
\item \ptthree, the transverse momentum of the third leading jet. 
  This variable tends to have higher values for signal than for background. 
\item \mjjmin, the lowest of the invariant masses of two jets, out of the three possible
  jet pairings.
  The simulation of this variable in \wpj\ production is discussed in Section~\ref{sec:syst}.
\item \lbpt, defined as for the \lpgefj\ channel, above. 
\item \djom, the difference in azimuthal angle between the 
  leading jet and the transverse momentum imbalance. 
  This variable provides additional discrimination between the MJ background and signal. 
  In MJ events the missing energy often originates from jet energy mismeasurement 
  and therefore tends to be directed opposite to the direction of the leading jet, whereas
  in \ttbar\ events the missing energy is generated by an escaping neutrino.
\end{itemize}
Jets that arise from gluon splitting are typical of \wpj\ and MJ production, 
and tend to have a low invariant mass and somewhat lower \pt\ than jets in \ttbar\ events.
Thus, low \mjjmin, \mjj, and \ktmin\ values are indicative of background.

The distributions of these variables in data and their modeling are shown in 
Fig.~\ref{fig:invarsW4} for \lpgefj\ events and in 
Fig.~\ref{fig:invarsW3} for \lptj\ events. 
The fractions of \ttbar\ signal, \wpj\ background and MJ background 
are taken from the results of the fit described in the next subsection. 
The number of events (\Nother) due to the other background processes, $Z$+jets, single top quark 
and diboson production, is fixed to the predicted value.

\begin{figure*}[htbp]
\includegraphics[width=0.32\linewidth]{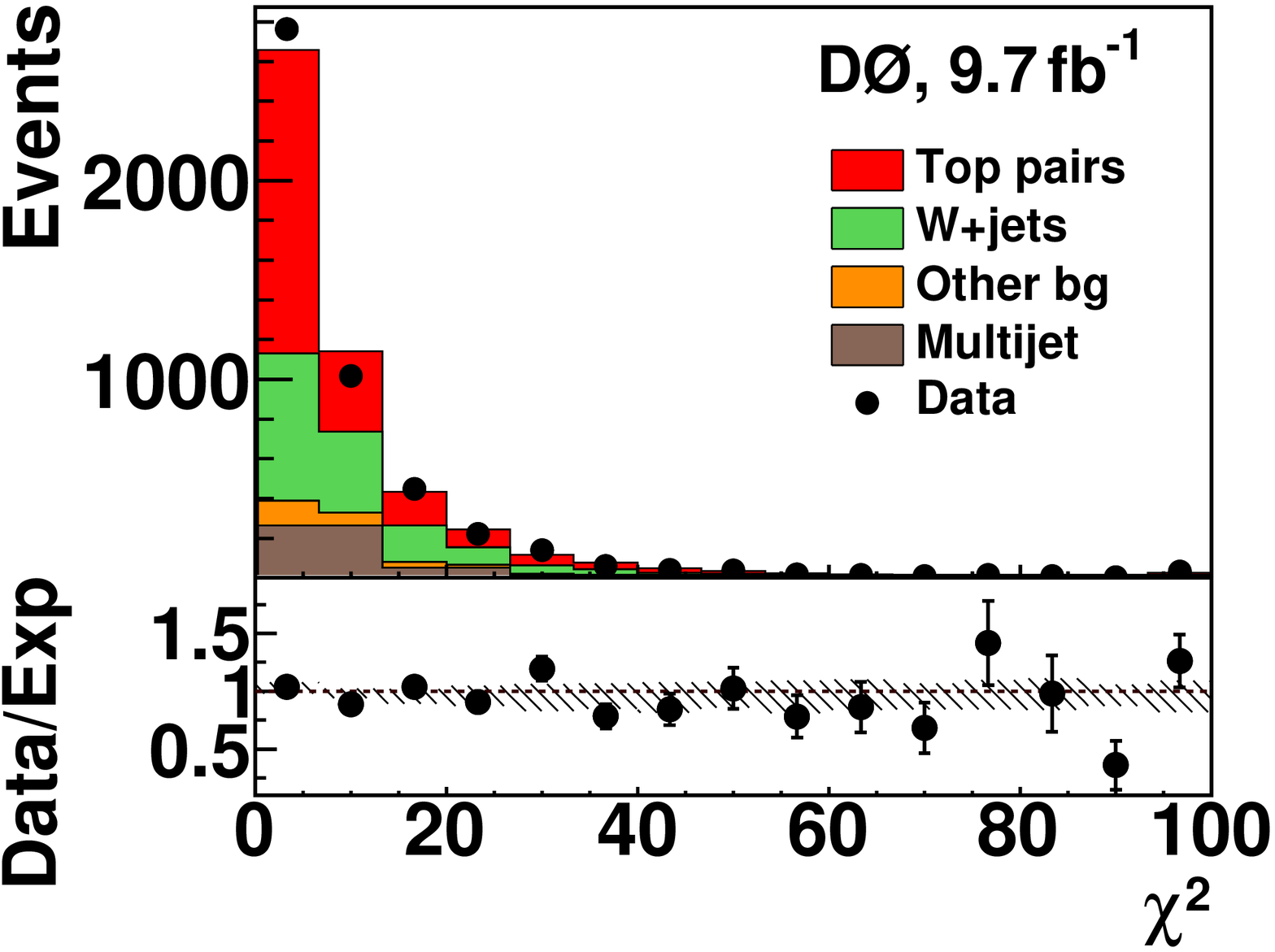}\hspace{0.04\linewidth}
\includegraphics[width=0.32\linewidth]{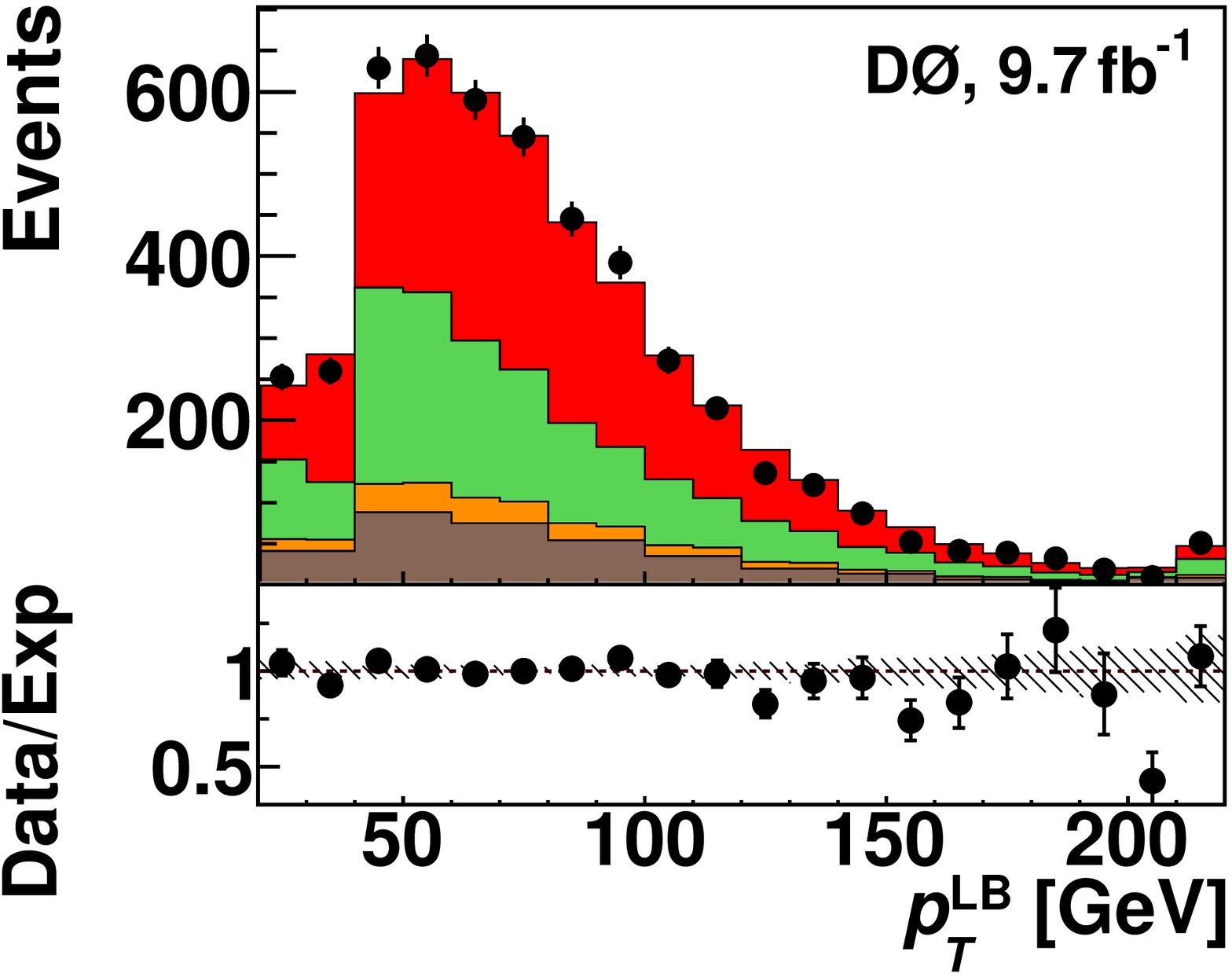}\\
\includegraphics[width=0.32\linewidth]{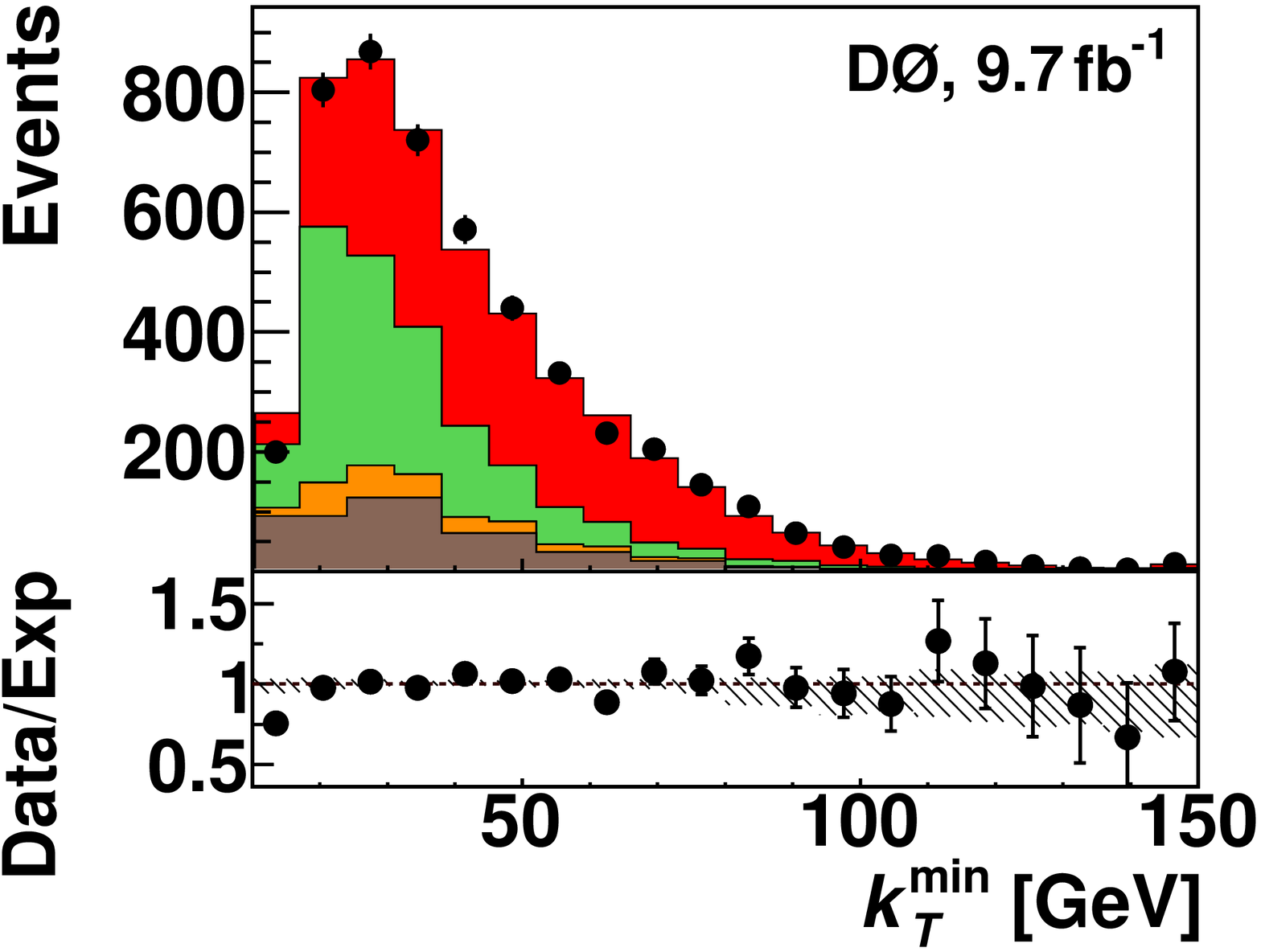}\hspace{0.04\linewidth}
\includegraphics[width=0.32\linewidth]{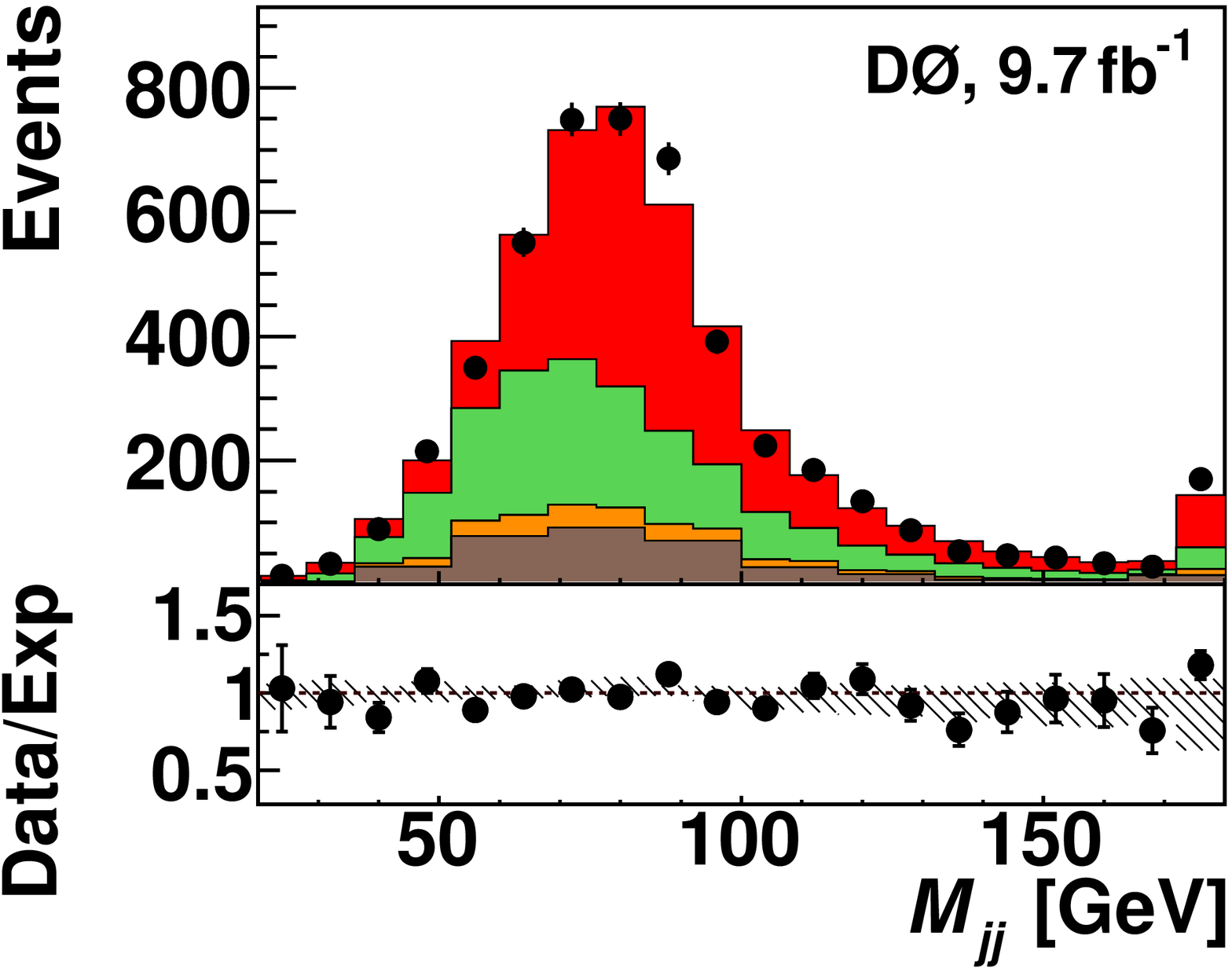}
\caption{
Input variables to the discriminant in the \lpgefj s sample (see Section~\ref{sec:disc} for definitions of variables).
Overflows are shown in the extreme bins.
The ratios between the data counts and the model expectations are shown in the lower panel of each figure.
The hashed area indicates the systematic uncertainties on the model expectations.
}
\label{fig:invarsW4}
\end{figure*}

\begin{figure*}[htbp]
\includegraphics[width=0.32\linewidth]{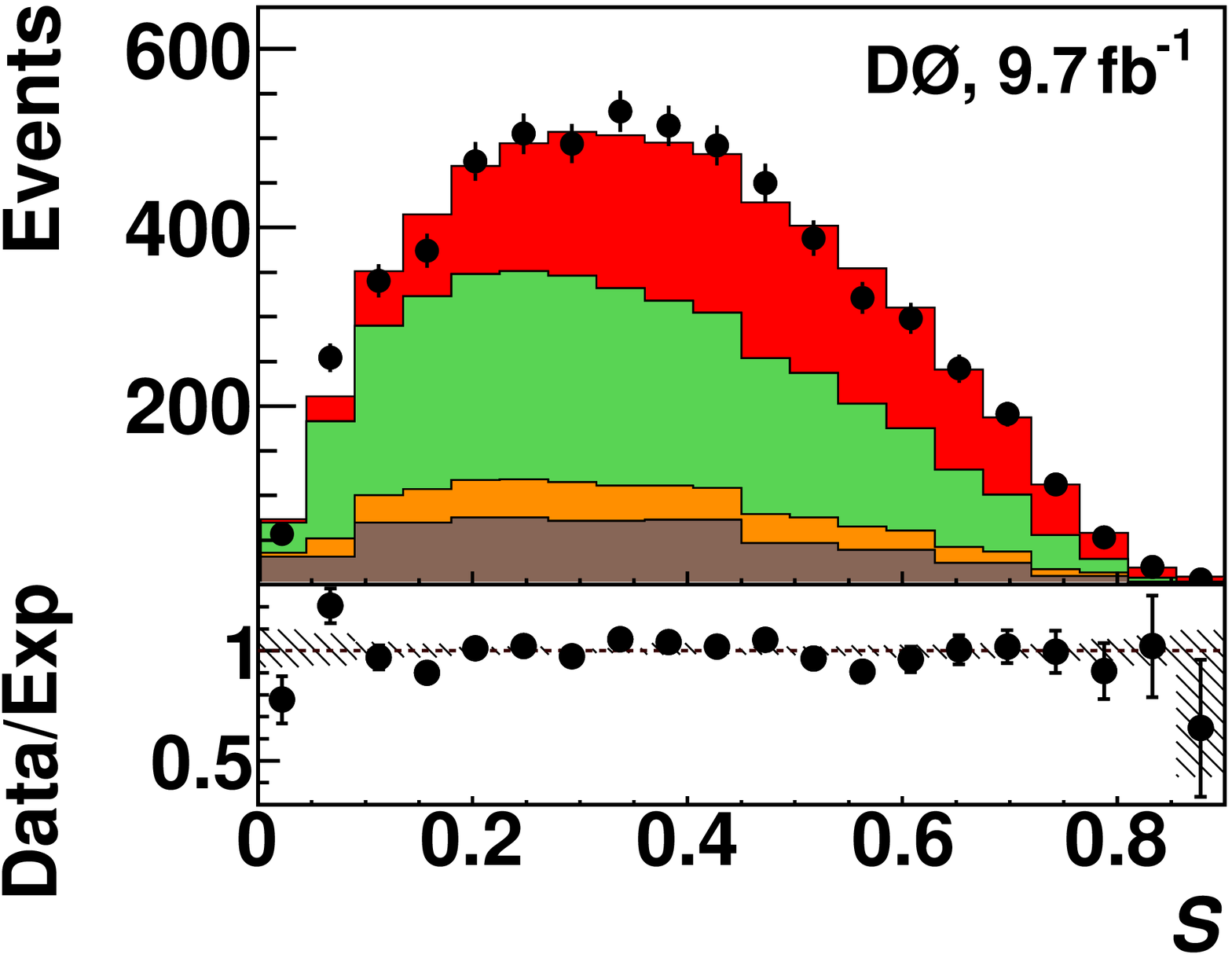}\hspace{0.04\linewidth}
\includegraphics[width=0.32\linewidth]{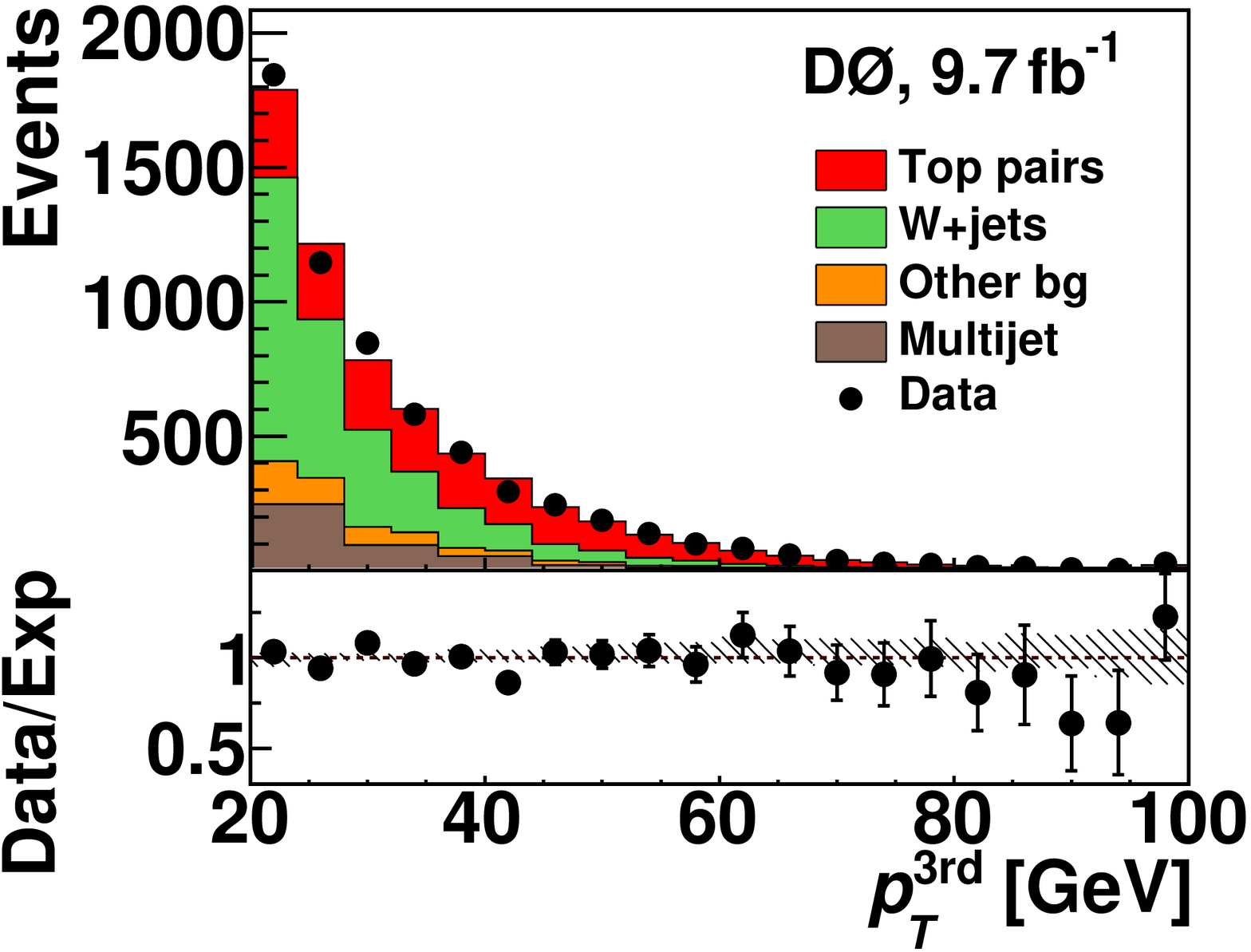}\\
\includegraphics[width=0.32\linewidth]{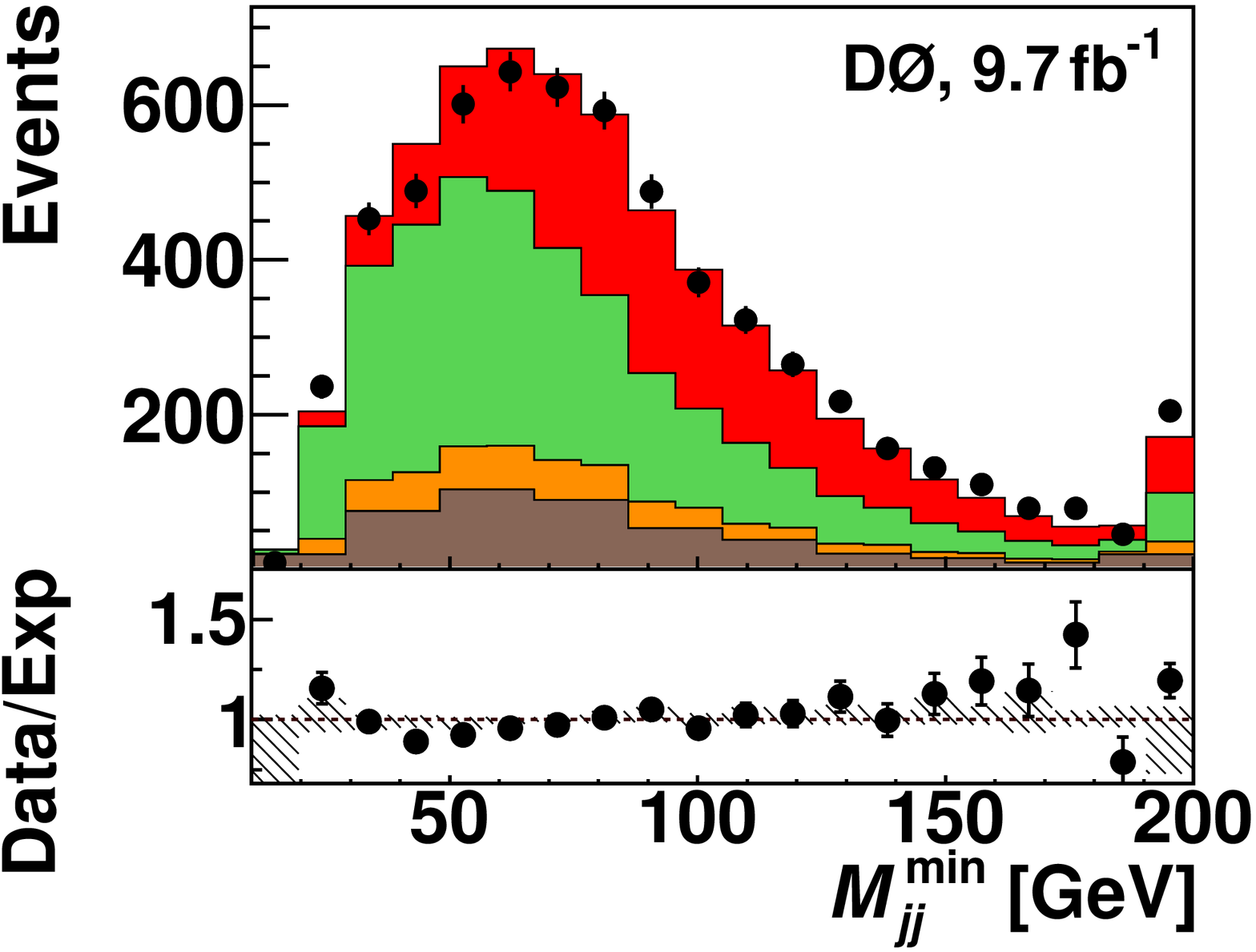}\hspace{0.04\linewidth}
\includegraphics[width=0.32\linewidth]{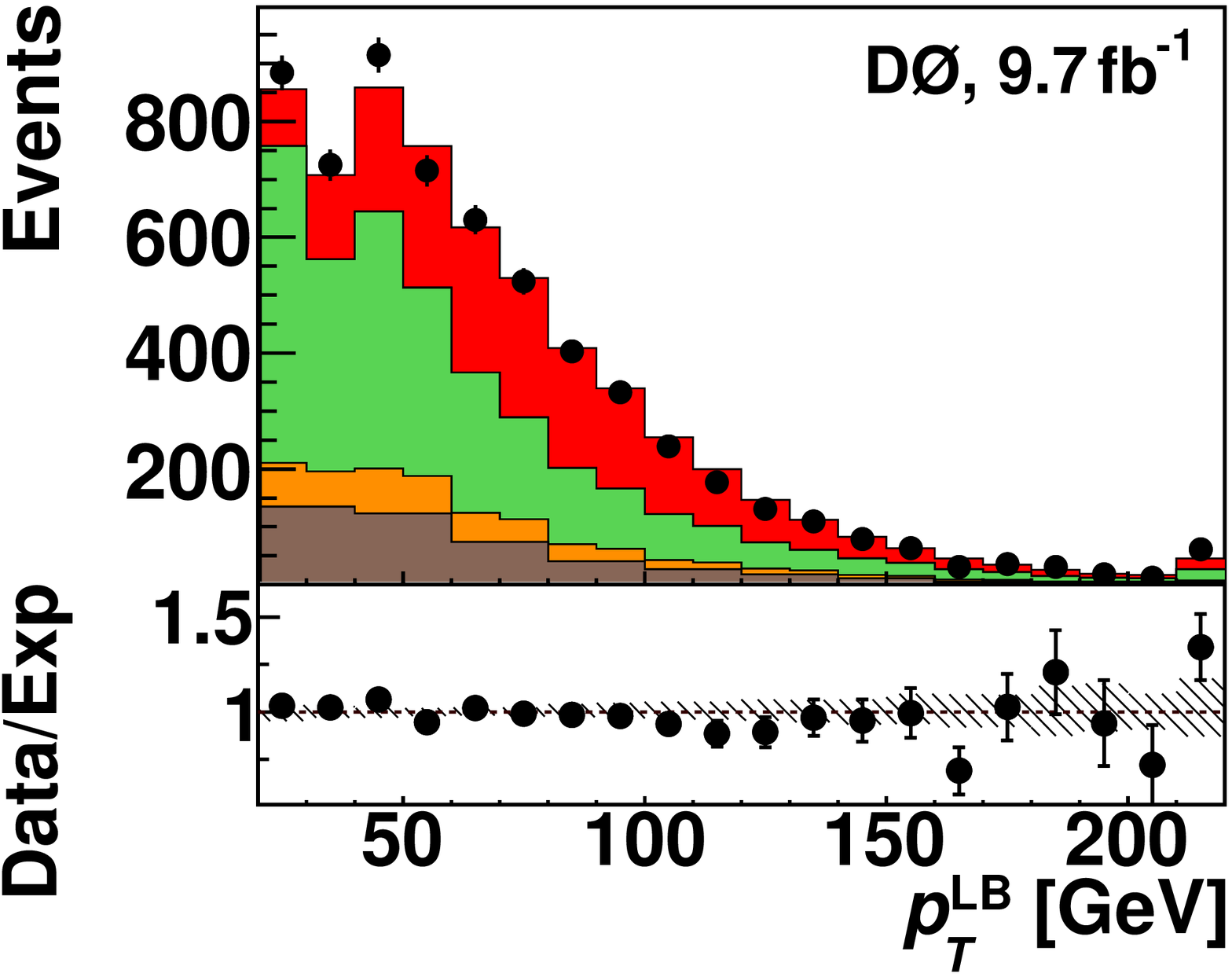}\\
\hspace{0.20\linewidth}\includegraphics[width=0.32\linewidth]{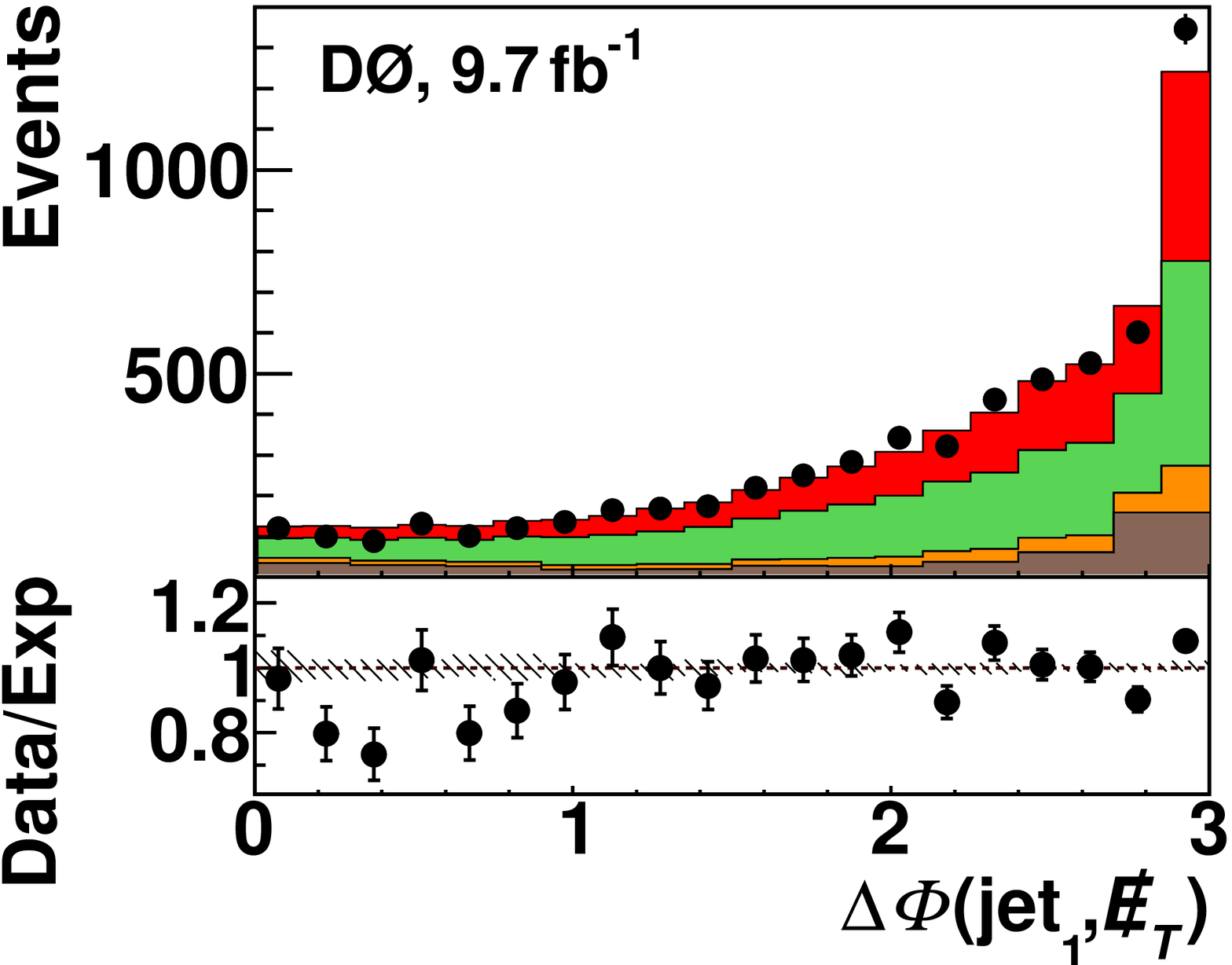}
\begin{picture}(132,5)(132,5)
\put ( 12,366){\subfloat[][]{\label{subfig:v3_a}}}
\put (150,378){\subfloat[][]{\label{subfig:v3_b}}}
\put ( 12,240){\subfloat[][]{\label{subfig:v3_c}}}
\put (198,240){\subfloat[][]{\label{subfig:v3_d}}}
\put ( 21,115){\subfloat[][]{\label{subfig:v3_e}}}
\end{picture}
\caption{
Input variables to the discriminant in the \lptj\ sample 
(see Section~\ref{sec:disc} for definitions of variables) for events with at least one $b$ tag.
Overflows are shown in the extreme bins.
The ratios between the data counts and the model expectations are shown in the lower panel of each figure.
The hashed area indicates the systematic uncertainties on the model expectations.
}
\label{fig:invarsW3}
\end{figure*}

%
%
\subsection{Maximum likelihood fit}
\label{sec:fit}

Selected events are categorized into six channels by the number of jets and $b$ tags.
The \lptj, zero-$b$-tag channel serves as a control region for the asymmetries of the \wpj\ background
while the other five ``signal'' channels are used in the maximum likelihood fit.
The number of selected \ttbar, \wpj, and MJ events in the five signal channels, i.e.,  the sample composition of the data sample,
and the reconstructed \afbl\ are extracted simultaneously using a maximum likelihood fit to the distributions of 
\disc\ and $\sgn(\qyl)$ (the $\sgn$ function is $1$ if its operand is positive and $-1$ otherwise)
across the five signal channels.
The distribution of the discriminant across all channels is shown in Fig.~\ref{fig:disc}.
The following four samples are used to construct the templates for the fit:
\begin{itemize}
\item simulated \ttbar\ signal events with $\qyl>0$,
\item simulated \ttbar\ signal events with $\qyl<0$,
\item simulated \wpj\ events,
\item a control data sample that has been enriched in MJ production
      by inverting the lepton isolation requirements~\cite{bib:D0xsect}.
\end{itemize}
The shape of the discriminant is the same for both signal templates.
Thus, their relative contribution is controlled by the $\sgn(\qyl)$ distribution,
which yields the fitted reconstruction\-/level asymmetry, 
after background subtraction. 

The normalization of the MJ background is determined using the observed number of events in the MJ-enriched
control sample and the probability of a jet to satisfy the lepton quality requirements~\cite{bib:D0xsect}.
The probability for jets to pass lepton quality requirements, particularly in the \mpj\ channel,
is dependent on \ptl.
We therefore split the MJ background template into six components, 
one for each lepton flavor and \ptl\ region.
The presence of signal in the MJ control sample (``signal contamination'') is accounted for
both in the likelihood and when calculating the relative
weights of the templates in the data model (\eg, in Figs.~\ref{fig:invarsW4} and~\ref{fig:invarsW3}).
To reduce statistical fluctuations in the \ptl-dependent measurement,
and in other fits of subsamples (see Section~\ref{sec:discussion}),
the number of bins of the MJ discriminant distributions is reduced by a factor of two,
and for the \hptl\ measurement, by a factor of three.

The results of this fit are given in Table~\ref{tab:afbl_fit}, where the measured \afbl\
values are from the fit done after the reweighting of the \wpj\ background described below.
In Fig.~\ref{fig:dy_chan}, the distributions of \qyl\ are taken from the simulated
samples, with the exception of the distribution for MJ production, which is modeled from the MJ-enriched 
control sample. 
The distribution for \wpj\ is shown after the reweighting described in the next section.

\begin{table}[htbp]
\caption{Predicted and measured \afbl\ values at reconstruction level, 
numbers of events estimated from signal and background sources,
and total numbers of events selected, excluding the three-jet zero-$b$-tag control data.
The quoted uncertainties on the measured values are statistical.
The \mcatnlo\ predictions are listed with their total uncertainties.
  \label{tab:afbl_fit}
}
\begin{ruledtabular}
\begin{tabular}{lcccc}
                & \multihead{4}{\ptl\ range, GeV} \\
\head{Quantity} & \head{\riptl} & \head{\rlptl} & \head{\rmptl} & \head{\rhptl} \\
\hline
Pred. \tablestrut$\afbl$, \% & $1.6 \pm 0.2$ & $1.2 \pm 0.5$ & $1.2 \pm 0.4$ & $2.3 \pm 0.3$\\
\hline
$\afbl$, \tablestrut\% &  $2.9\pm2.1$ & $-1.2\pm4.1$ & $3.0 \pm 3.2$ & $7.2\pm3.6$ \\
\Nw  & $4445 \pm 68$ & $1609 \pm 40$ & $1842 \pm 45$ & $1008 \pm 41$ \\
\Nmj & $969 \pm 23$ & $325 \pm 13$ & $309 \pm 14$ & $333 \pm 14$\\
\Nother & 787 & 271 & 319 & 197\\
\Ntt & $4746\pm64$ & $1341 \pm 38$ & $1951 \pm 43$ & $1438 \pm 38$\\
\hline
\Nsel \smallstrut& $10947$ & $3548$ & $4422$ & $2977$ \\
\end{tabular}
\end{ruledtabular}
\end{table}

\begin{figure*}[htbp]
\centering
\includegraphics[scale=.72]{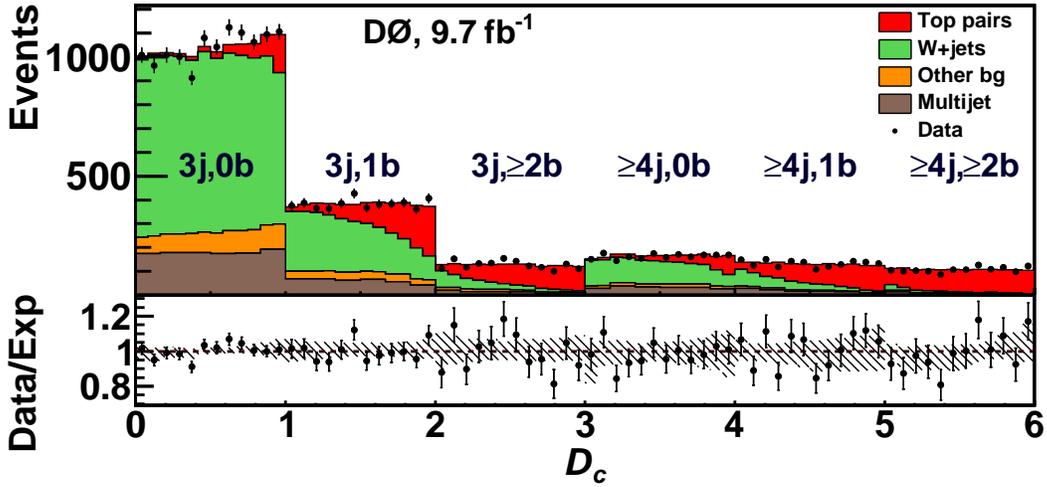}
\vspace{-0.2cm}
\caption{
The discriminants for all channels, concatenated into a single variable \dc.
Each unit of \dc\ corresponds to a channel, as labeled in the plot.
The region $\dc<1$ is not used in the fit for sample composition and \afbl.
The ratio of the data counts and the model expectation is shown below.
The hashed area indicates the systematic uncertainties on the model expectations.
}
\label{fig:disc}
\end{figure*}

\begin{figure*}[htbp]
\centering
\includegraphics[scale=.32]{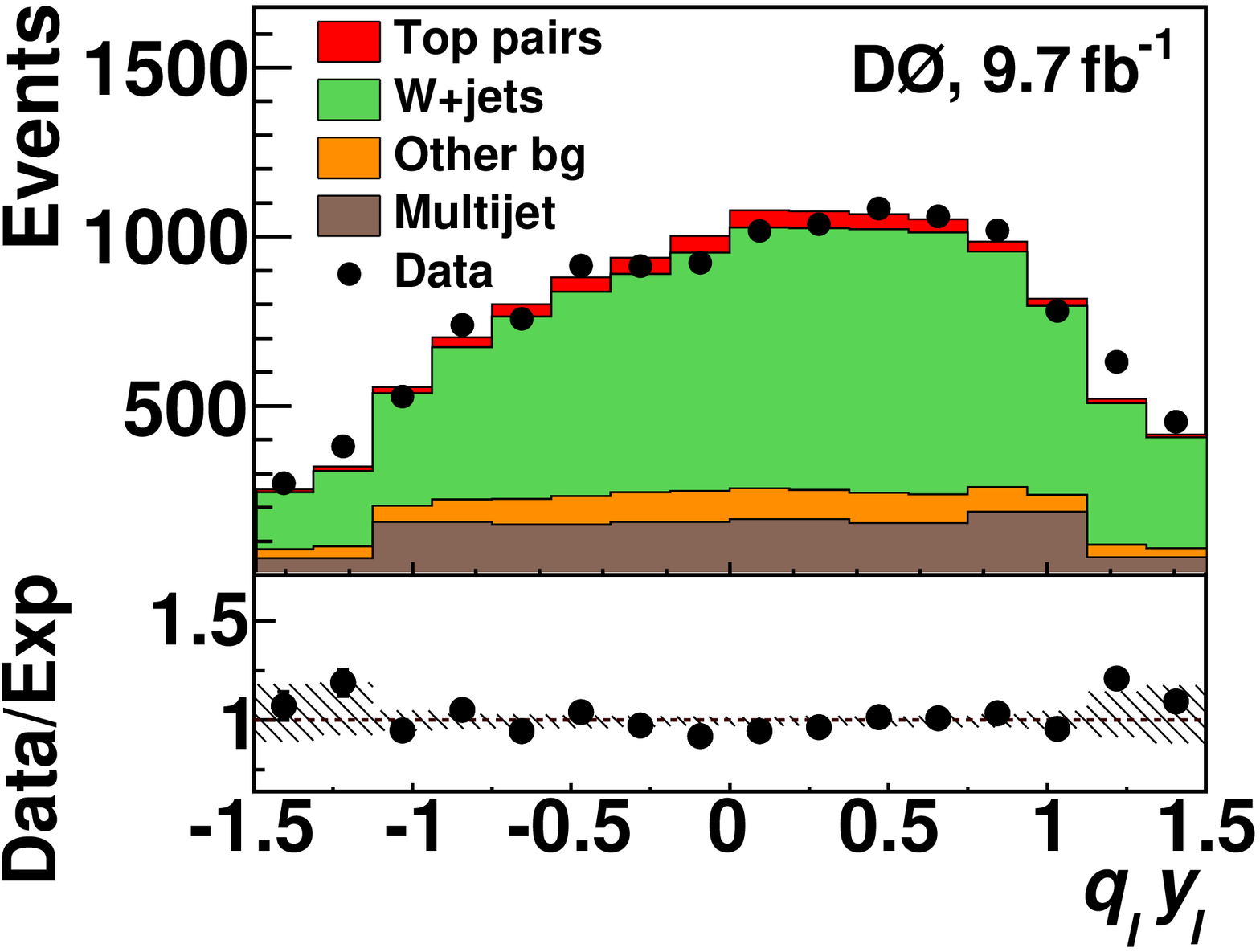} \hspace{0.04\linewidth}
\includegraphics[scale=.32]{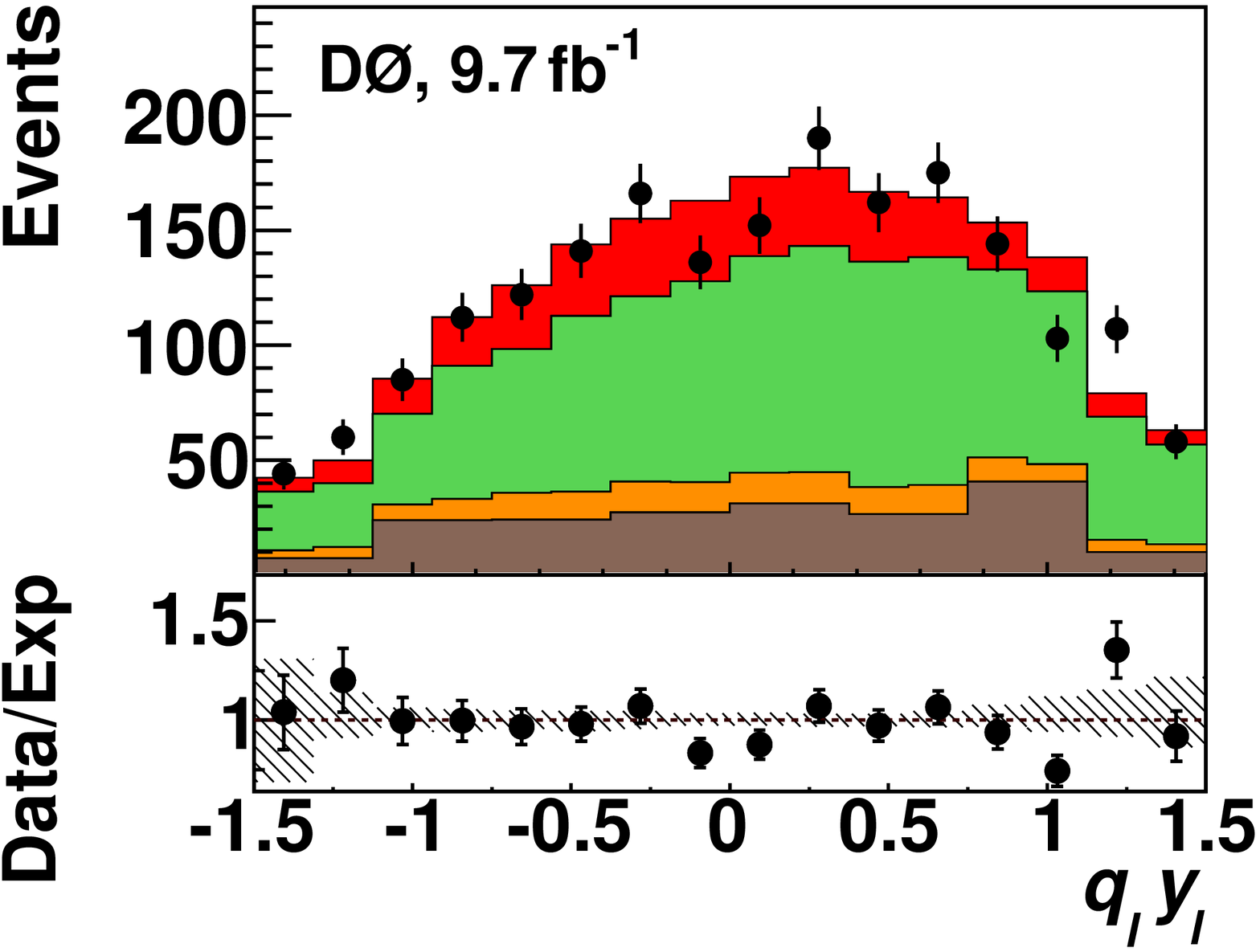} \\
\includegraphics[scale=.32]{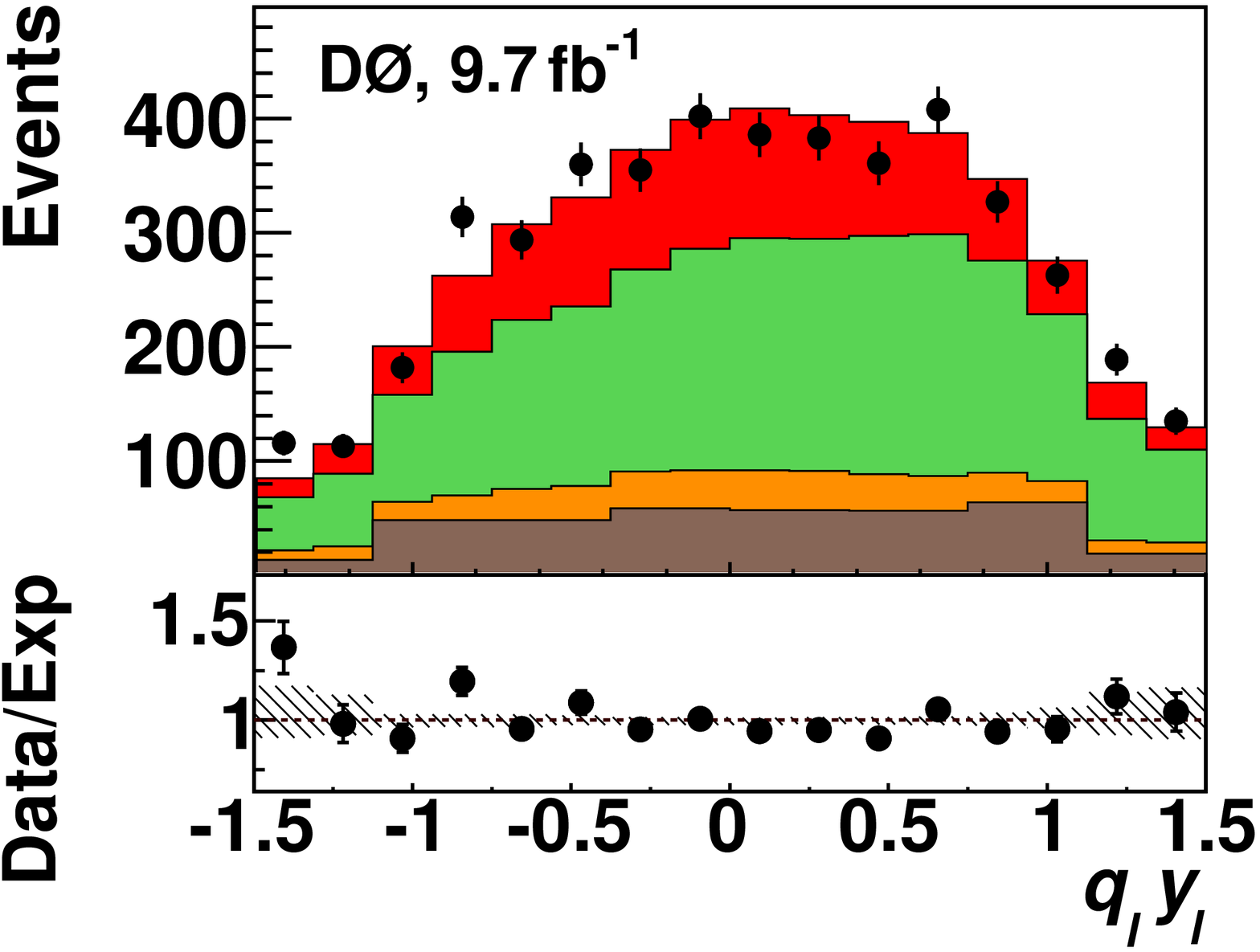} \hspace{0.04\linewidth}
\includegraphics[scale=.32]{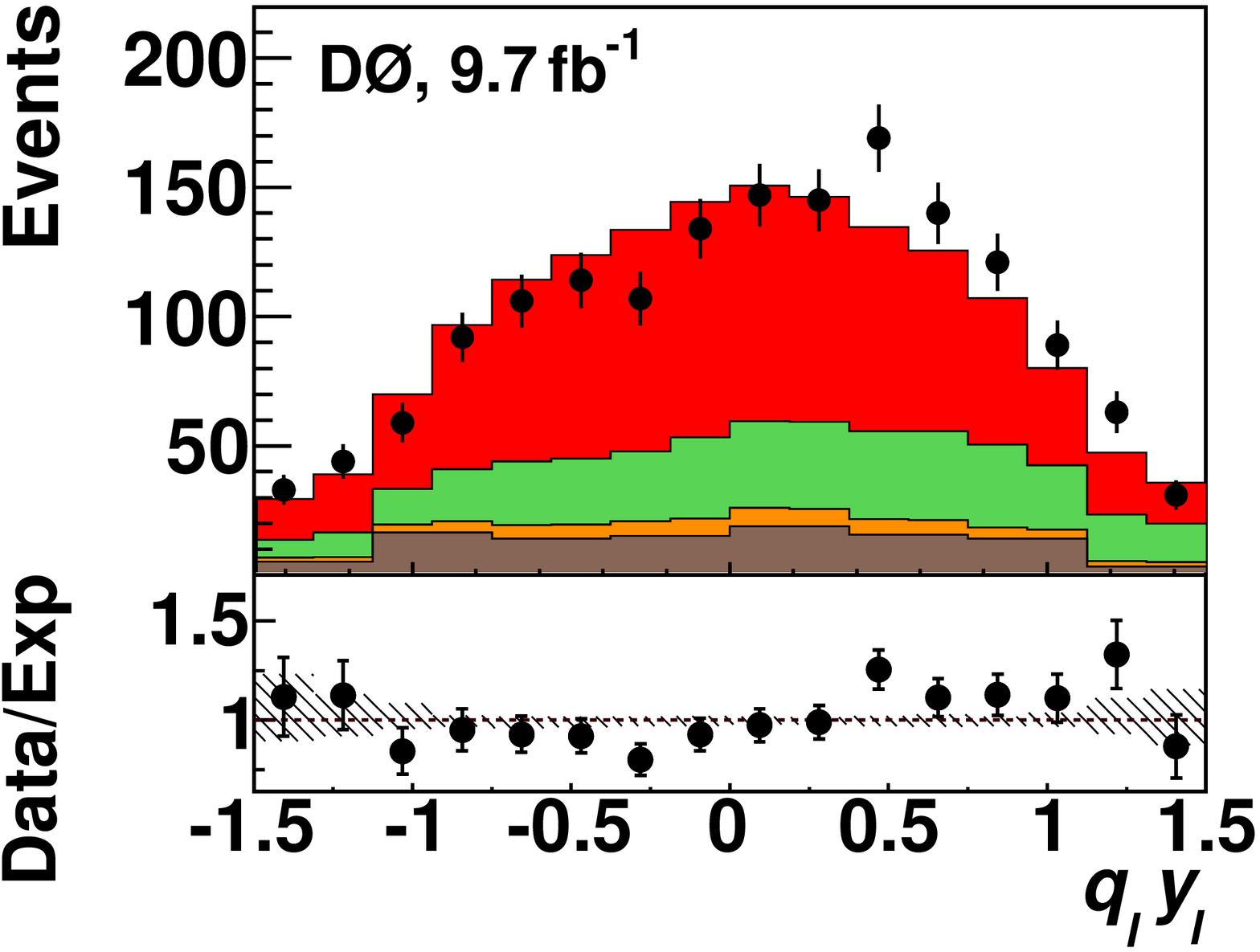} \\
\includegraphics[scale=.32]{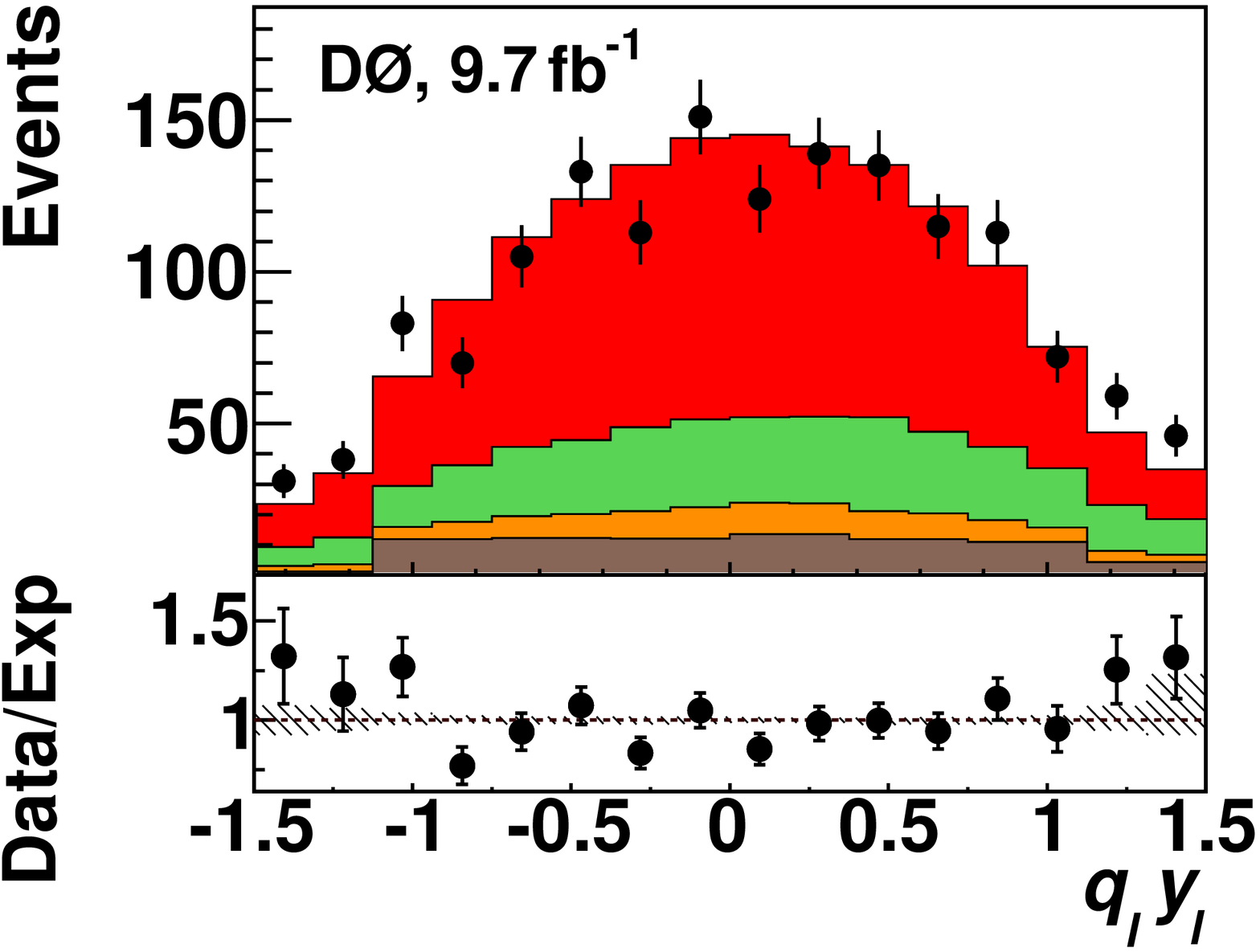} \hspace{0.04\linewidth}
\includegraphics[scale=.32]{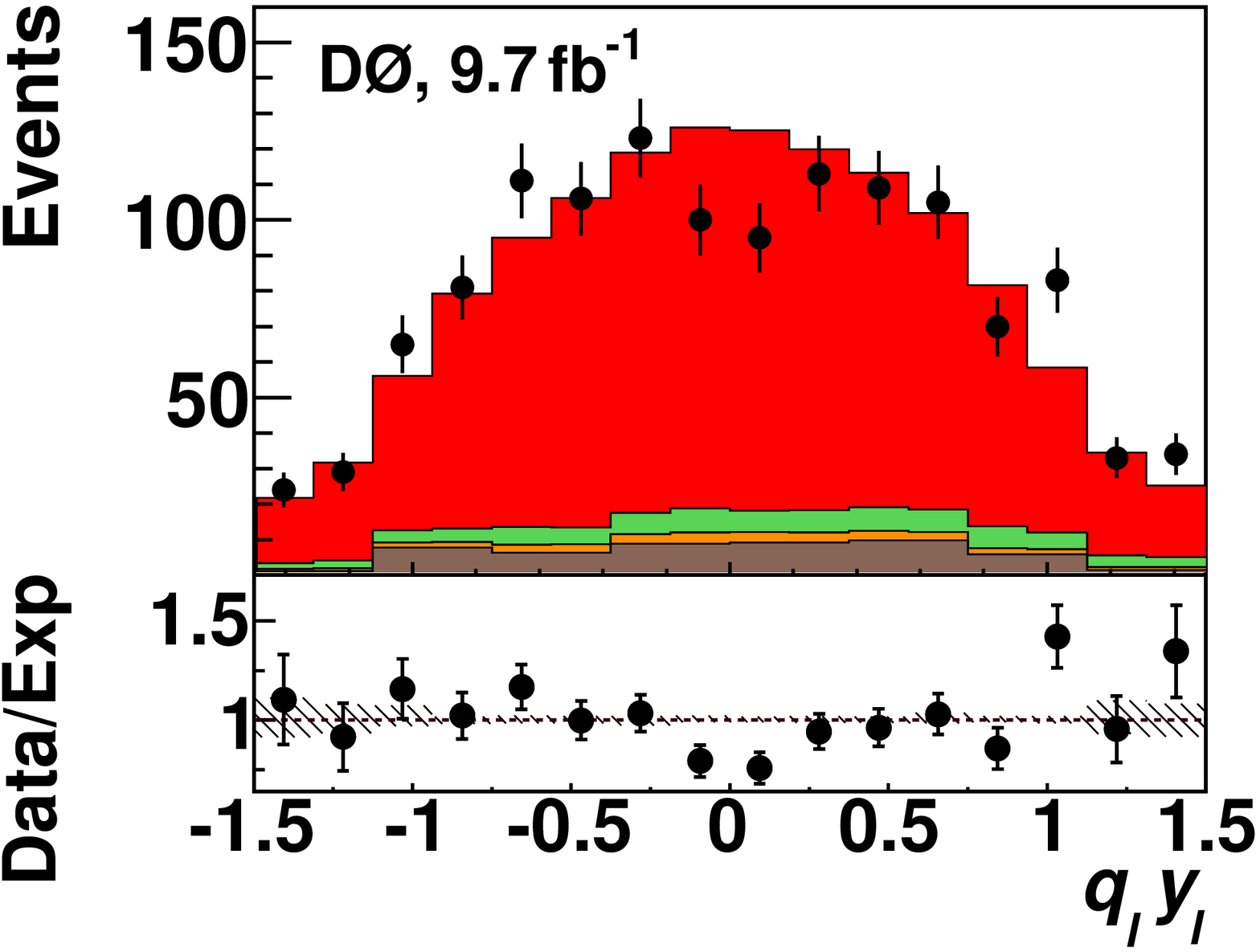} 
\begin{picture}(132,5)(132,5)
\put (158,420){\subfloat[][]{\label{subfig:dy_a}}}
\put (366,432){\subfloat[][]{\label{subfig:dy_b}}}
\put (158,292){\subfloat[][]{\label{subfig:dy_c}}}
\put (366,292){\subfloat[][]{\label{subfig:dy_d}}}
\put (158,152){\subfloat[][]{\label{subfig:dy_e}}}
\put (366,152){\subfloat[][]{\label{subfig:dy_f}}}
\end{picture}
\vspace{-0.4cm}
\caption{
The \qyl\ distribution for: (a) \lptj s and 0 $b$ tags, (b)  \lpgefj s and 0 $b$ tags,
(c) \lptj s and 1 $b$ tag, (d)  \lpgefj s and 1 $b$ tag,
(e) \lptj s and \getb, and (f)  \lpgefj s and \getb.
The ratio between the data counts and the model expectation is shown below each plot.
The hashed area indicates the systematic uncertainties on the model expectations.
}
\label{fig:dy_chan}
\end{figure*}

%
%
\subsection{\boldmath Reweighting the simulated \wpj\ background}
\label{sec:wcalib}

Since both the \wpj\ background and the \ttbar\ signal contribute to the \afbl\ of the selected data,
accurate modeling of \afbl\ in \wpj\ production is required for the measurement of \afbl\ in \ttbar\ events.
The asymmetry has been measured precisely for inclusive \wbprod~\cite{bib:winc}.
However, there are notable differences between inclusive \wbprod\ and the production of a $W$ boson 
in association with jets, which constitutes the main background in this analysis.
In particular, inclusive \wbprod\ is dominated by collisions between valence $u$ and $\bar{d}$ 
(or $d$ and $\bar{u}$) quarks.
As the average momentum carried by $u$ quarks is higher than that carried by $d$ quarks,
the $W^+$ bosons are preferentially boosted in the direction of the incoming proton.
The boost of the $W$ bosons leads to positive \afbl\ in inclusive \wbprod,
which dominates over the negative contribution to \afbl\ due to the $V-A$ nature of \wdbos\ decay.
But in the \wpj\ events that pass the selection criteria of this analysis and contribute to the background,
the rate and properties of the events that originate from interactions between valence quarks are different.
In these events, the production of multiple jets reduces the boost of the $W$ bosons relative to their
boost in inclusive $W$ production, leading to negative \afbl.
Only 20\%--40\% of the \wpj\ background originates from interactions between valence quarks
and the rest originates from quark--gluon interactions, which produce a positive \afbl.

We compare the simulated \afbl\ to the control data sample with three jets and zero $b$ tags, 
which is dominated by \wpj\ background and is not used for measuring \afbl\ in \ttbar\ events.
The composition of the three-jet, zero-$b$-tag control sample cannot be determined reliably by applying
the technique of the previous section solely within the control sample itself (see Fig.~\ref{fig:disc}).
Using the normalizations of the \ttbar\ signal and MJ background from the fit of Section~\ref{sec:fit}, we
predict their contributions in the control sample.
We find that the control sample is dominated by \wpj\ background, 
with about 75\% of events from \wpj\ production,
12\% from MJ production, 9\% from other backgrounds and 4\% from \ttbar\ production.

The differential asymmetry $\afbl\left(\absyl\right)$ is constrained by
continuity to be zero at $\absyl=0$, hence at first order it is proportional to \absyl.
Figure~\ref{fig:wpj_weight} shows $\acr\left(\absyl\right)$, 
where \acr\ is the \afbl\ of the \wpj\ background in the control region,
and fits of $\acr\left(\absyl\right)$ to a line that passes through the origin
for the \wpj\ simulation and for data. The control data is shown after subtraction of
the estimated contributions from \ttbar, MJ, and other-background production.

\begin{figure*}[htbp]
\centering
\includegraphics[scale=.31]{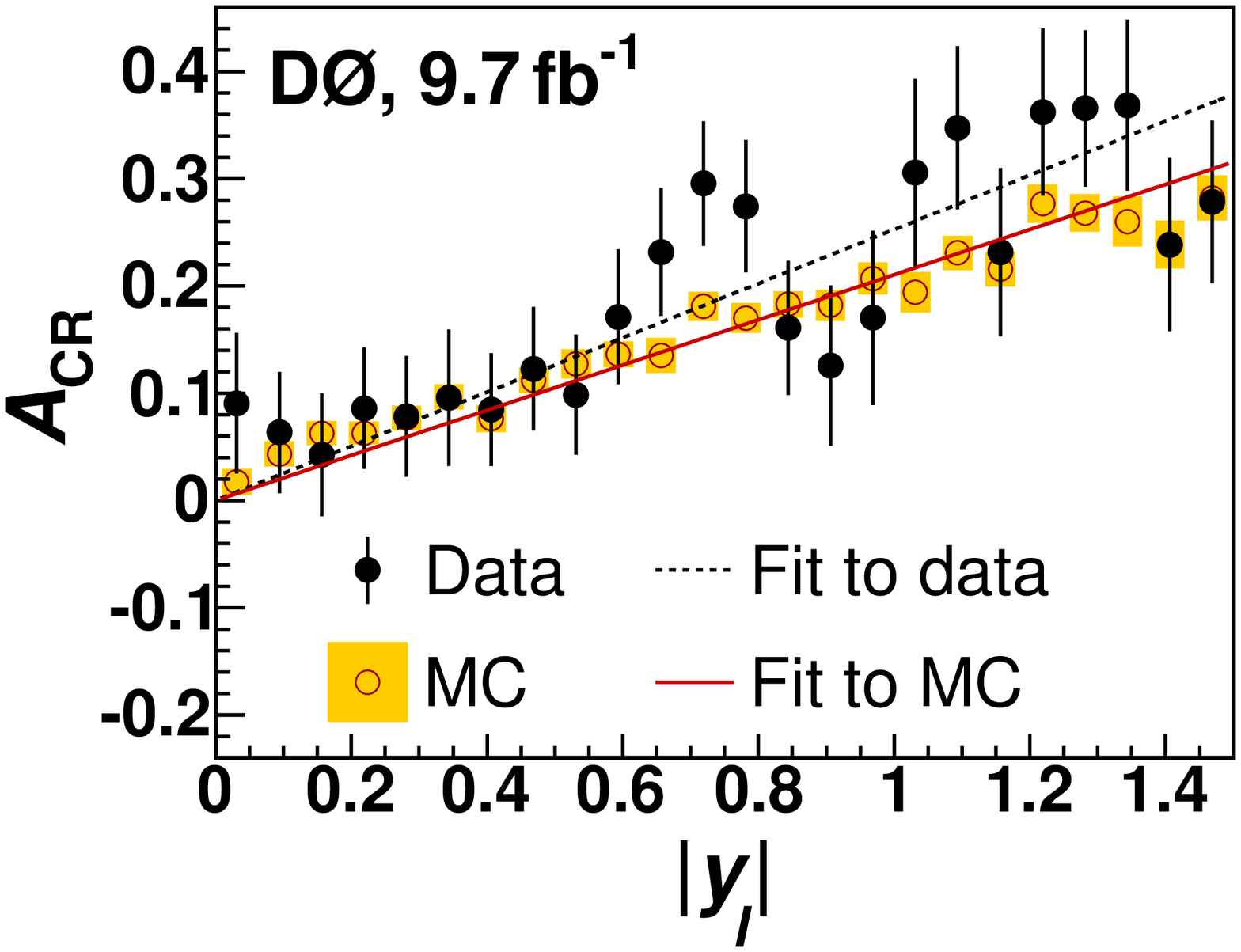} \hspace{0.04\linewidth}
\includegraphics[scale=.31]{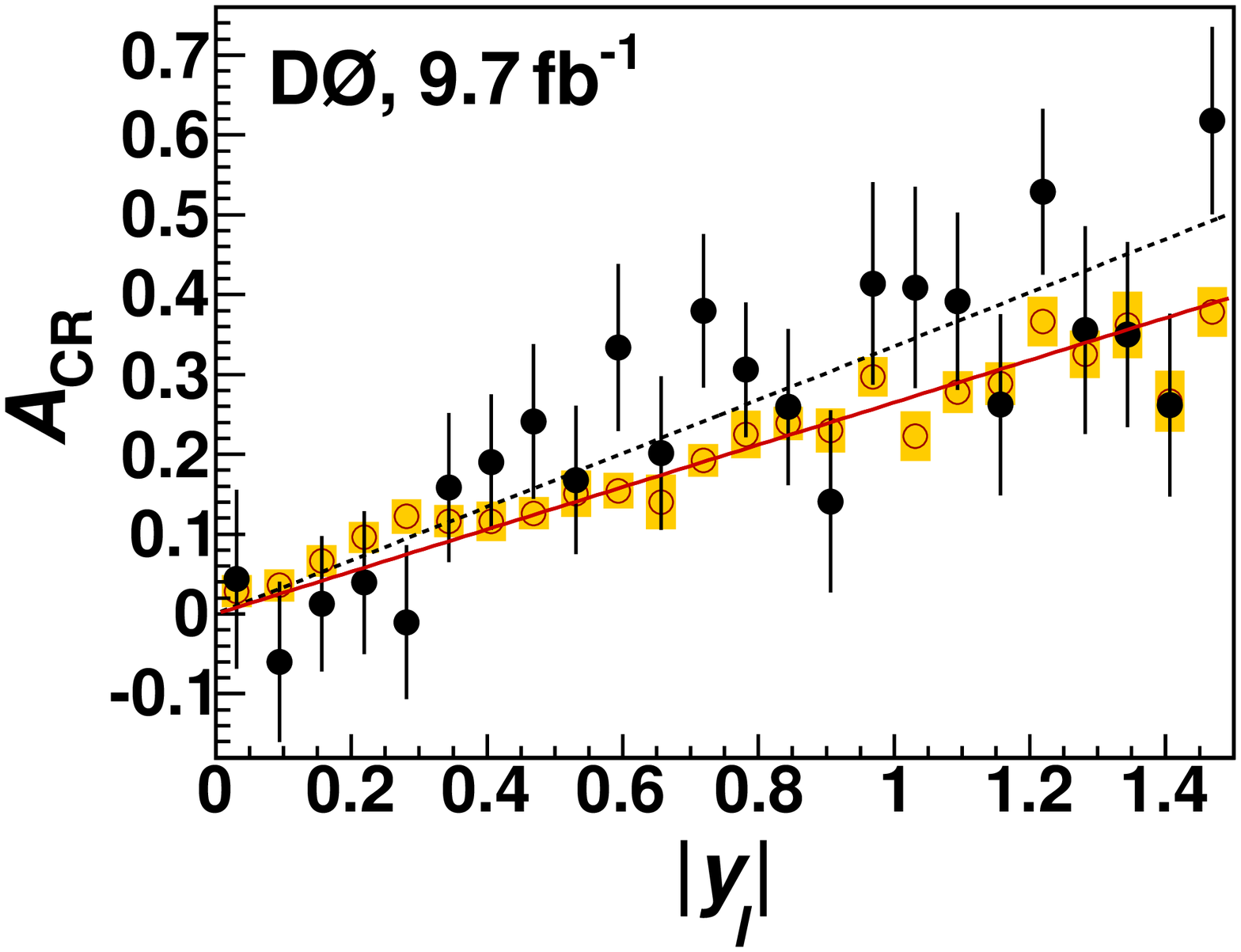}\\
\includegraphics[scale=.31]{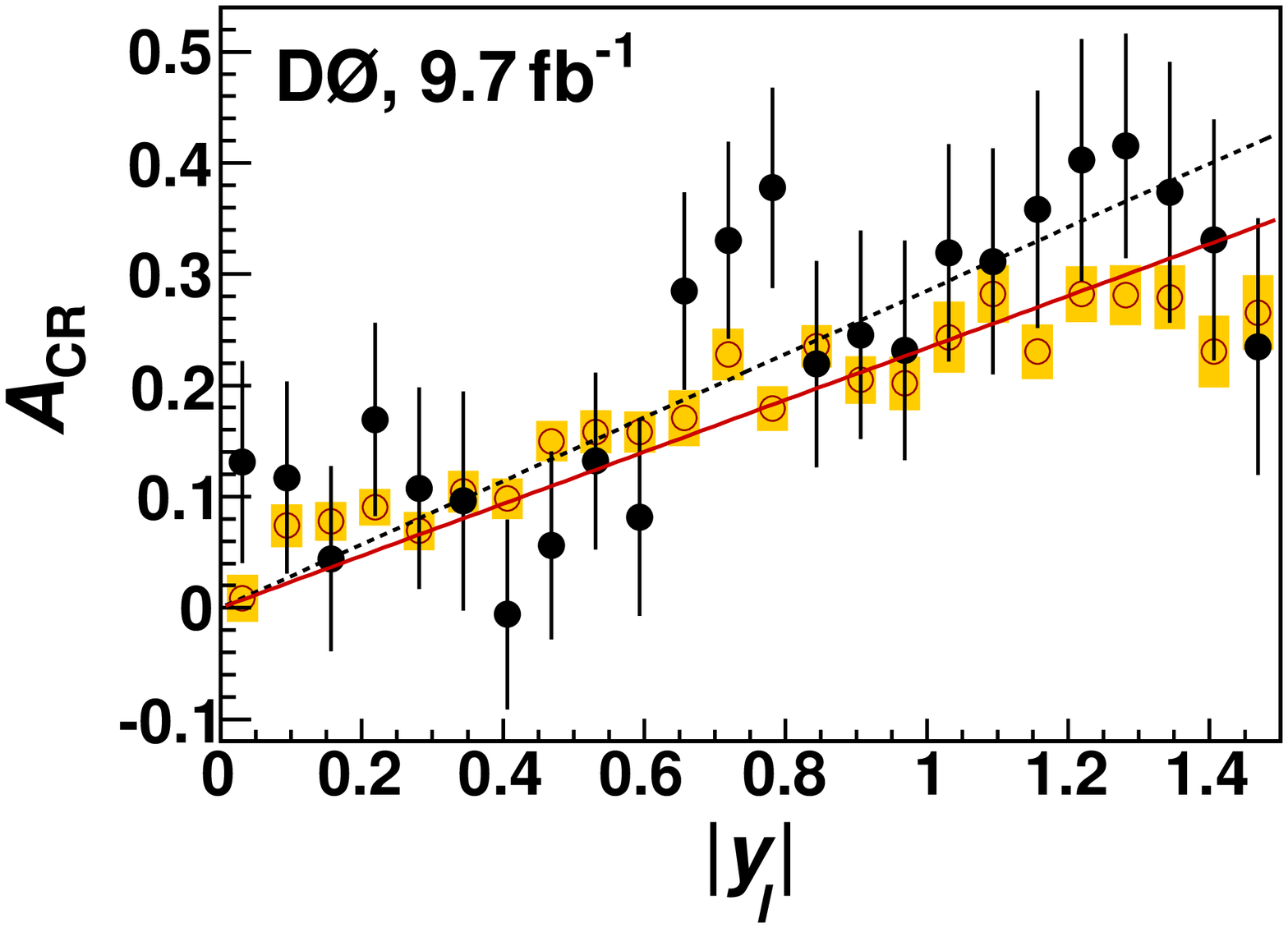}\hspace{0.04\linewidth}
\includegraphics[scale=.31]{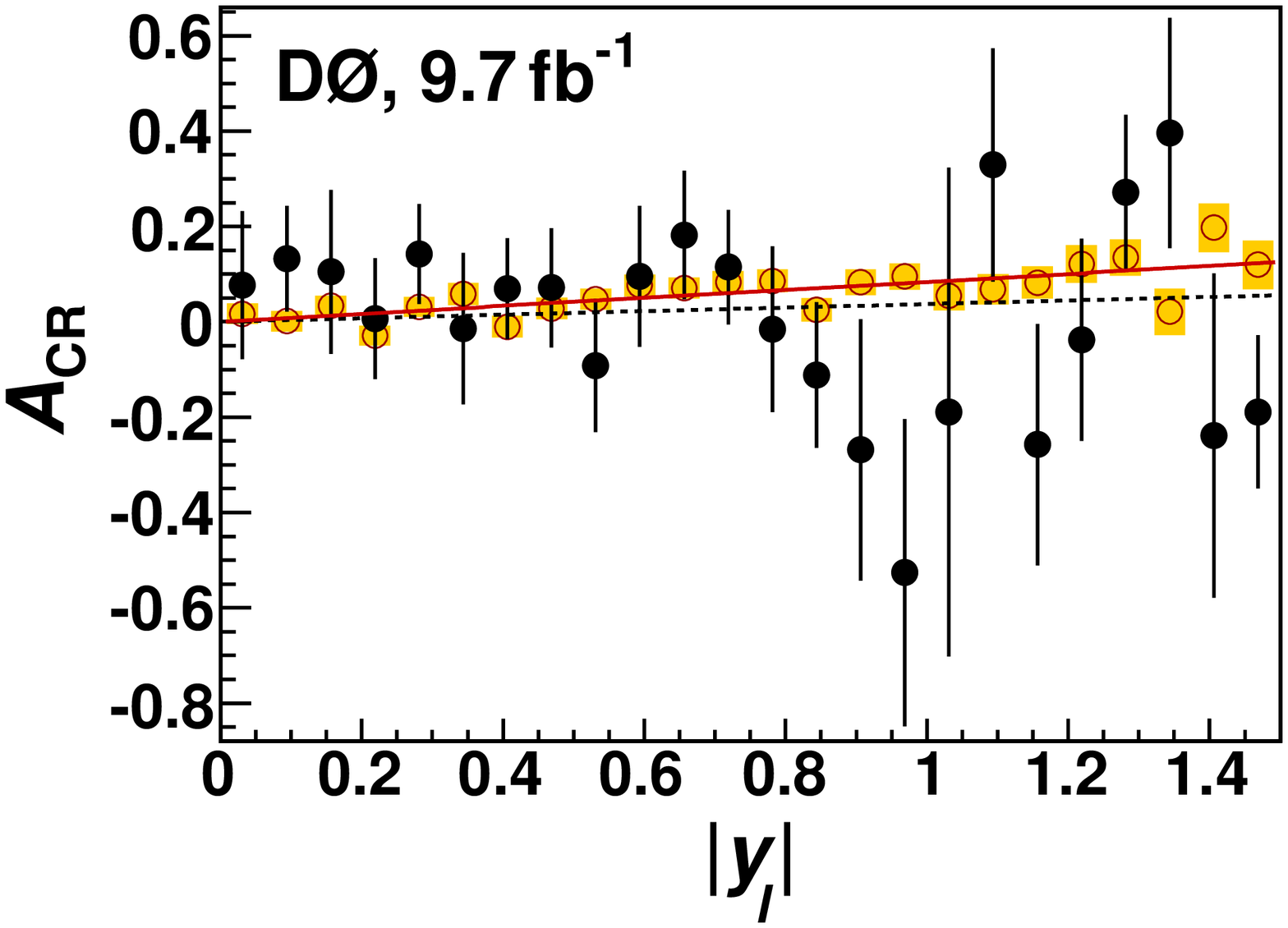}
\begin{picture}(350,0)(0,0)
\put ( 25,254){\subfloat[][]{\label{subfig:wc_a}}}
\put (227,254){\subfloat[][]{\label{subfig:wc_b}}}
\put ( 26,120){\subfloat[][]{\label{subfig:wc_c}}}
\put (290,133){\subfloat[][]{\label{subfig:wc_d}}}
\end{picture}
\vspace{-0.6cm}
\caption{
The asymmetry of leptons from \wpj\ production as a function of \absyl\ for (a) the inclusive sample,
(b) \lptl, (c) \mptl, and (d) \hptl.
The points show the data from the control region,
after the subtraction of the non-\wpj\ contributions, and the dashed line 
shows a fit to the functional form $y=ax$. 
The empty circles and solid line show the nominal \wpj\ simulation 
and its fit to the same functional form.
The error bars and shaded regions indicate the statistical uncertainties
on the data and simulation.
}
\label{fig:wpj_weight}
\end{figure*}

We weight each event of the simulated \wpj\ background, 
regardless of its jet and $b$-tag multiplicities, using the function
\begin{equation}
w = 1 + \alpha q_{l}y_{l},    \label{eq:alpha}
\end{equation}
choosing $\alpha$ so that the simulated slope of $\acr\left(\absyl\right)$
(shown in Fig.~\ref{fig:wpj_weight} for $\alpha=0$) agrees with the observed slope.
Thus, we rely on the MC generators to describe the dependence of the \wpj\ \afbl\ 
on the number of jets and the number of $b$ tags, 
and the resulting channel-to-channel differences in this asymmetry.
The statistical uncertainty on $\alpha$ is taken from the statistical uncertainties 
on the slopes of the fits to both data and MC.
The resulting differences \dacr\ between the \acr\ before and after the reweighting
are larger than expected from the PDF uncertainties on \acr\ (see Table~\ref{tab:calib}).
This raises the possibility that this tension is not entirely due to the choice of PDFs
and leads us to assign the entire effect of the reweighting as a systematic uncertainty.

\begin{table}[htbp]
\caption{Parameters of the \qyl\ reweighting of the \wpj\ background, effects on \acr, and PDF uncertainties. 
The first row lists the parameter $\alpha$ of the \qyl\ reweighting with its statistical uncertainty.
The second row lists the effect of the reweighting on \acr.
The next two rows list the up and down uncertainties on \acr\ due to PDFs.
\label{tab:calib}
}
\begin{ruledtabular}
\begin{tabular}{lcccc}
                & \multihead{4}{\ptl\ range, GeV} \\
\head{Quantity} & \head{\riptl} & \head{\rlptl} & \head{\rmptl} & \head{\rhptl}  \\
\hline
$\alpha$, \%  & $4.5 \pm 1.8$ & $7.9 \pm 2.7$ & $5.7 \pm 2.4$ & $-6.6 \pm 4.3$\\
\dacr, \%  & $2.7 \pm 1.0$ & $4.7 \pm 1.6$ & $3.3 \pm 1.4$ & $-3.9 \pm 2.6$\\
\spwup, \% & $1.0$ & $0.5$ & $1.2$ & $0.8$ \\
\spwdown, \% & $1.7$ & $1.6$ & $1.7$ & $1.8$ \\
\end{tabular}
\end{ruledtabular}
\end{table}

The \qyl\ reweighting, using the control data, reduces the PDF uncertainties by at least a factor of three in each \ptl\ range. 
However, for each \ptl\ bin the uncertainties on the \afbl\ of the \wpj\ background from the 
reweighting procedure are more than twice the size of the simulated PDF uncertainties
evaluated without the reweighting procedure.

The production\-/level measurement is affected by this \qyl\ reweighting through
the \qyl\ distribution of the subtracted \wpj\ background. 
The reconstruction\-/level measurement is affected by this reweighting through \dacr.
The effect of the reweighting on the \wpj\ \afbl\ in each signal channel is within $0.3\%$ of \dacr.
%
%
\section{\boldmath Unfolding the asymmetries}
\label{sec:unfold}

The inclusive \afbl\ is unfolded separately in each channel, 
and the measured \afbl\ values are then combined to form the inclusive measurement.
Due to the excellent angular resolution for leptons, migrations in rapidity are negligible,
and unfolding the \afbl\ reduces to correcting for acceptance effects. 
The inverse of the simulated selection efficiency is taken as a weight for each bin 
in \qyl\ to correct for acceptance effects. 
These corrections therefore assume the SM as modeled in \mcatnlo.
We restrict the selection to $\absyl<1.5$ to avoid the region of low acceptance, 
and compute the weights in 48 bins as in the previous \afbl\ measurement~\cite{bib:ourPRD}.

For the differential \afbl\ measurement, we define for each of the 48 \qyl\ bins a vector $\vec{r}$ of 
observed counts in the three \ptl\ bins.
The observed counts are affected by the migration of $\approx 10\%$ of the events
over the bin boundaries in \ptl. 
The expectation value of $\vec{r}$ is 
$\langle\vec{r}\hspace{1pt}\rangle=\accmat\migmat\vec{p}$, where \accmat\ is the acceptance matrix, 
\migmat\ is the $3\times3$ migration matrix
and $\vec{p}$ is the vector of production\-/level event counts. 
The acceptance matrix is a $3\times3$ diagonal matrix with the three acceptance probabilities 
embedded in its diagonal.
The vector of the unfolded production\-/level counts that best estimates the vector $\vec{p}$ is
$\vec{u}=\accmat^{-1}\migmat^{-1}\vec{r}$.
With a nearly diagonal migration matrix and only three bins, the above matrix inversion
yields stable solutions.

We evaluate the statistical uncertainty of the unfolded \afbl\
from each channel using an ensemble of pseudo-datasets that match
the sample composition in data, with the signal simulated according to \mcatnlo. 
The pseudo-datasets are simulated using Poisson fluctuations both on the selected
sample and on the MJ control sample. 
The statistical uncertainties on \afbl\ for each channel 
and the weight of each channel in the combined measurement are listed in Table~\ref{tab:stat_weight}.

\begin{table}[htbp]
\caption{Statistical uncertainty ($\sigma$) on the measured \afbl\ and weight for each channel (where applicable).
The weight for each channel is proportional to $\sigma^{-2}$. \label{tab:stat_weight}
}
\begin{ruledtabular}
\begin{tabular}{lcc}
 \head{Channel} &  \head{$\sigma$} & \head{Weight}\\
\hline
\lptj s, 0 $b$ tags &  $24\%$  & n/a    \\
\lptj s, 1 $b$ tag  &  $6.8\%$ & $0.11$ \\
\lptj s, \getb      &  $4.7\%$ & $0.24$  \\
\hline
\lpgefj s, 0 $b$ tags &  $13.9\%$ & n/a  \\
\lpgefj s, 1 $b$ tag  &  $4.7\%$ & $0.24$ \\
\lpgefj s, \getb      &  $3.6\%$ & $0.41$ \\
\end{tabular}
\end{ruledtabular}
\end{table}

Since the \lptj, zero-$b$-tag channel is used to tune the modeling of the \wpj\
background, it cannot be used to extract the signal \afbl.
We also do not use the \lpgefj, zero-$b$-tag channel for the unfolded result,
due to its low purity and the large uncertainty on \afbl.
The weighted average of the four remaining $b$-tagged channels gives our 
combined value for \afbl. 

The lepton-based asymmetries unfolded to the production level are 
summarized in Table~\ref{tab:lafb} and shown in Fig.~\ref{fig:ptl_dep}. 
The results are compared to \mcatnlo-based predictions.

\begin{table}[htbp]
\caption{Predicted and observed production\-/level asymmetries.   
  The first uncertainty on the measured \afbl\ is statistical and the second systematic.
  The statistical uncertainties on the MC predictions are less than $0.1\%$,
  while the scale and PDF uncertainties are estimated to be $<1\%$.
  \label{tab:lafb}
}
\begin{ruledtabular}
\begin{tabular}{l  c  c  }
&  \multihead{2}{\afbl, \%} \\
\head{\ptl\ range, GeV}& \head{Data} & \head{MC@NLO} \\
\hline
Inclusive \tablestrut& $4.2 \pm 2.3^{+1.7}_{-2.0}$ & 2.0 \\[0.05cm]
\hline
\rlptl \tablestrut&  $-0.3\pm4.1 \pm 3.6$ & 1.6 \\[0.1cm]
\rmptl &  $4.8 \pm 3.5 ^{+2.2}_{-2.1} $ & 2.3 \\[0.1cm]
\rhptl &  $9.3 \pm 3.7 ^{+2.3}_{-2.7} $ & 3.1\\
\end{tabular}
\end{ruledtabular}
\end{table}

\begin{figure}[htbp]
\begin{center}
\includegraphics[width=0.8\linewidth]{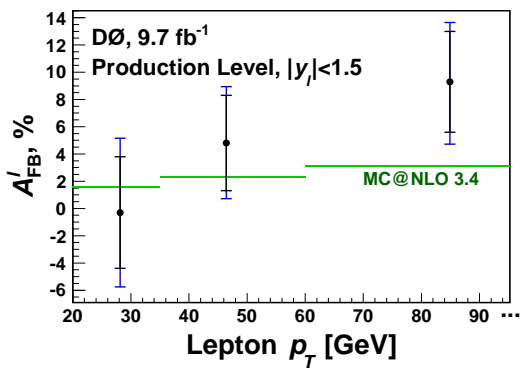}
\end{center}
\vspace{-0.6cm}
\caption{
Predicted and observed production\-/level asymmetries as a function of lepton transverse momentum. 
The last bin extends beyond the edge of the plot and has no upper boundary.
Statistical uncertainties are indicated by the inner,
and the total uncertainties by the outer error bars.
}
\label{fig:ptl_dep}
\end{figure}

We also measure the differential asymmetry as a function of \absyl\ by applying the same
procedure that is used for the inclusive asymmetry to the \qyl\ bins contained in each \absyl\ range.
The measured differential asymmetries are listed in Table~\ref{tab:absyl_prod} and shown in Fig.~\ref{fig:absyl_dep}. 

\begin{table}[htbp]
\caption{Predicted and observed production-level asymmetries as a function of \absyl. 
  The first uncertainty on the measured values is statistical and the second is systematic. 
  The statistical uncertainties on the MC predictions are less than $0.1\%$,
  while the scale and PDF uncertainties are estimated to be $<1\%$.
  \label{tab:absyl_prod}
}
\begin{ruledtabular}
\begin{tabular}{l c c}
&  \multihead{2}{\afbl, \%} \\
\head{\absyl\ range} & \head{Data} & \head{MC@NLO}  \\
\hline
\hphantom{$.123$}$0$ -- $0.125$ \tablestrut & \,\,\,$0.5 \pm 6.1^{+0.8}_{-0.7} $ & 0.2 \\
$0.125$ -- $0.375$              \tablestrut & \,   $0.5  \pm 4.4^{+1.3}_{-1.8} $ & 0.9 \\
$0.375$ -- $0.625$              \tablestrut & \,   $2.6  \pm 4.7^{+1.7}_{-1.5} $ & 1.8 \\
$0.625$ -- $1$                  \tablestrut & \,   $1.9  \pm 4.6^{+2.0}_{-2.3} $ & 2.7 \\
\hphantom{$.123$}$1$ -- $1.5$   \tablestrut &     $13.2  \pm 6.5^{+2.6}_{-3.0} $ & 3.7 \\
\end{tabular}
\end{ruledtabular}
\end{table}

\begin{figure}[htbp]
\begin{center}
\includegraphics[width=0.8\linewidth]{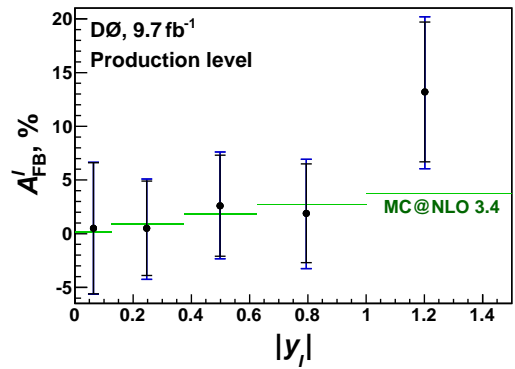}
\end{center}
\vspace{-0.6cm}
\caption{
Predicted and observed production\-/level asymmetries as a function of absolute lepton rapidity.
Statistical uncertainties are indicated by the inner,
and the total uncertainties by the outer error bars.
}
\label{fig:absyl_dep}
\end{figure}

%
%
\section {Systematic Uncertainties}
\label{sec:syst}

We consider several sources of systematic uncertainty. 
For most sources, we vary the modeling according to the evaluated uncertainty in the relevant parameters
of the model, repeat
the entire analysis and propagate the effect to the final result. 
This accounts for the correlations between the channels
and between the various steps of the analysis, such as the maximal likelihood fit,
the fit for $\alpha$, and the unfolding.
Some sources are quantified using more specialized procedures, as described below.
Systematic uncertainties from different sources are added in quadrature to yield the 
total systematic uncertainty.
Table~\ref{tab:sysAfbl} lists the systematic uncertainties on the predicted 
reconstruction\-/level \afbl\ (as listed in Tables~\ref{tab:preds} and~\ref{tab:afbl_fit}),
on the measured reconstruction\-/level \afbl, and on the measured production\-/level \afbl.
The systematic sources are classified into the following categories:
\begin{description}[\breaklabel]
\item[Jet reconstruction (reco)]
This source includes the jet reconstruction and identification efficiencies, 
as well as the efficiency of the vertex confirmation described in Section~\ref{sec:selection}. 
The simulated efficiencies are calibrated on dijet data. 
Additional \ppbar\ collisions within the same bunch crossing can yield additional jets, and their
modeling is also included in this category.
The rate of additional \ppbar\
collisions is derived from the number of reconstructed vertices per event.

\item[Jet energy measurement] 
The jet energy scale (JES) is measured using dijet and photon+jet samples~\cite{bib:JES}.
The simulated jet energy resolution (JER) is calibrated using $Z+$jet data.
Their uncertainties are propagated to the measured asymmetry.

\item[Signal modeling]
As discussed in Section~\ref{sec:preds}, the SM predicts a negative asymmetry
for events with additional final state gluons (and hence with larger \ttpt).
Thus, event selection introduces a bias on the measured asymmetry,
in particular making it sensitive to the jet multiplicity. The inclusion of \lptj\ events in the
analysis reduces this correlation. To evaluate the size of this systematic effect we vary the amount of 
initial-state radiation (ISR) within an uncertainty range established
from a measurement of ISR rates~\cite{bib:Zisr}. 

Forward-backward differences in the amount of additional radiation can also affect the measurement through
\ttpt, which is correlated with the acceptance~\cite{bib:ourPRD}.  
These differences are controlled by the simulated color coherence of the partonic showers~\cite{bib:ourPRD}.
QCD predicts that parton showers in angular order are more likely, 
while the simulation enforces strict angular ordering~\cite{bib:ttpt}.
NLO event generators calculate the first QCD emission analytically, 
reducing the reliance on the modeling of the parton showers.
To quantify this uncertainty, we consider the possibility that the dependence of \afbl\ on \ttpt\
is 25\% smaller than in \mcatnlo, a possibility motivated by the studies of Ref.~\cite{bib:ttpt}.

The \ttpt\ distribution was studied in Refs.~\cite{bib:CDFdep,bib:ttpt_others} and found to be well modeled.
This is in contrast to the limitations of the \DZ-detector simulation, which result
in poor modeling of this distribution~\cite{bib:ourPRD}. We consider the possibility that this mismodeling
also affects \afbl\ by reweighting the simulated events as a function of the reconstructed \ttpt\ so
that the \ttpt\ distribution agrees with data.

The mass of the top quark was varied from its value in the nominal simulation of $172.5\GeV$
according to the latest measurement~\cite{bib:mtop}.
To quantify additional systematic uncertainties due to the modeling of signal, 
we repeat the analysis using signal events simulated with \alpgen\ combined with \pythia.
As the box diagram is not included in \alpgen, 
\alpgen\ predictions are missing the largest contribution to the \ttbar\ asymmetry~\cite{bib:k_and_r}.
Furthermore, this missing contribution peaks at low \ttpt, where acceptance is low, 
making the acceptance predicted by \alpgen\ unrealistic.
Therefore we use the acceptance predicted by \mcatnlo\ instead of the one predicted by \alpgen\ in evaluating
this uncertainty.

The uncertainties on the production\-/level inclusive \afbl\ due to the top quark mass, the choice of MC generator,
and the overall amount of ISR are similar.
The systematic uncertainties due to the forward-backward differences 
in the additional radiation and due to \ttpt\ reconstruction have negligible effect on the inclusive \afbl.

\item[\boldmath $b$ tagging]
The $b$-tagging efficiency and mistagging probability, which are determined from dijet data with at least 
one muon identified within a jet, affect the division of events between 0, 1, and $\ge$ 2 $b$-tag subsamples. 
Due to this division of channels, the analysis is now more sensitive to systematic variations on $b$-tagging 
than the previous measurement~\cite{bib:ourPRD}.

\item[Background (Bg) subtraction]
The subtracted amounts of \wpj\ and MJ background are varied within their fitted uncertainties. 
Uncertainties on the normalization and shape of the MJ background arise from 
the uncertainties on the lepton selection rates, which are used to evaluate the MJ background.
An uncertainty of $20\%$ is assigned to the rate of \Wcc\ and \Wbb\ production.

\item[Background (Bg) modeling]
The \qyl\ reweighting of the \wpj\ background is varied using $\alpha$ values of zero and twice
the nominal $\alpha$ (see Table~\ref{tab:calib}).
The effect of increased MJ production at large \absyl\ is considered by reweighting
the MJ \qyl\ distribution to better match the data in the \lptj, zero-$b$-tag control region. 
The possibly underestimated muon background with mismeasured high transverse momentum
described in Section~\ref{sec:selection} peaks in that region, and 
an excess of data events in that region is seen in some channels.
We also consider a similar increase in \wpj\ production at large \absyl.

We account for the marginal agreement of the dijet invariant mass (see Fig.~\ref{subfig:v3_c})
and related observables between data enriched in \wpj\ production and the \alpgen\ simulation of such 
data~\cite{bib:wjets_reweight} 
by reweighting the simulated \mjjmin\ distribution of the \wpj\ background to match data in the \lptj, 0-$b$-tag control region.
This improves the modeling of \mjjmin\ in all channels, supporting the attribution of this small mismodeling
to the modeling of \wpj\ production. 
A small mismodeling of \djom\ is indicated in Fig.~\ref{subfig:v3_e}, but its effect on the discriminant 
is far smaller than that of the region around $\mjjmin=50\GeV$.

\item[Parton distribution functions]
Each of the error eigenvectors of the set of PDFs is varied up and down, and the effects
are added in quadrature.
We also consider an uncertainty due to the choice of PDFs, which we evaluate
using the nominal PDFs of the CTEQ6L1~\cite{bib:cteq} and MRST2003~\cite{bib:MRST} sets.
The MRST2003 set is chosen since its $u$, $d$, $s$, and $g$ PDFs differ significantly from those of
the CTEQ6L1 set for values of Bjorken $x$ above $0.01$, which are the most relevant to this analysis.
\end{description}

\begin{table}[htbp]
  \caption{
    Systematic uncertainties on \afbl.     
    Uncertainties smaller than 0.1\% are omitted.
    \label{tab:sysAfbl}
  }
  \begin{ruledtabular}
    \begin{tabular}{lccc}
      &  \multicolumn{3}{c}{Absolute uncertainty, \%} \\
      &  \multicolumn{2}{c}{Reconstruction level} & Prod. level \\
      Source & Prediction & Measurement & Measurement\\
      \hline \tablestrut 
      Jet reco        &   $-0.1$   &     \emph{--}   &   \emph{--}     \\
      JES/JER         &   $+0.1$   & ${+0.1}/{-0.3}$ & ${+0.2}/{-0.3}$ \\
      Signal modeling & \emph{--}  & ${-0.2}$        & ${+0.6}/{-0.4}$ \\
      $b$ tagging     & $\pm{0.1}$ & ${+0.5}/{-0.8}$ & ${+0.8}/{-1.1}$ \\
      Bg subtraction  &  n/a       & ${+0.1}/{-0.3}$ & ${+0.1}/{-0.3}$ \\
      Bg modeling     &  n/a       & ${+1.4}/{-1.5}$ & ${+1.3}/{-1.5}$ \\
      PDFs            & \emph{--}  & ${+0.3}/{-0.2}$ & ${+0.1}/{-0.2}$ \\
      \hline
      Total           & $\pm0.1$   & ${+1.5}/{-1.7}$ & ${+1.7}/{-2.0}$\\
    \end{tabular}
  \end{ruledtabular}
\end{table}

%
%
\section{Discussion}
\label{sec:discussion}
Using a dataset corresponding to an integrated luminosity of $9.7\ifb$, 
we measure the production\-/level inclusive \afbl\ to be 
$\big(4.2\pm2.3\stat^{+1.7}_{-2.0}\syst\big)$\%.
The previously published value, which was measured using a subset of this dataset 
corresponding to 5.4\ifb~\cite{bib:ourPRD}, is $\afbl=\left(15.2\pm3.8\right)\%$. 
In this section we further compare these measurements and 
discuss the reasons for the difference between their results.
Unlike the previous measurement, 
the current result is in good agreement with the \mcatnlo\ prediction for the production\-/level of 
$\afbl=2.0\%$, which has a statistical uncertainty of less than $0.1\%$.
The statistical uncertainty on the measured \afbl\ is reduced by a factor of $\approx1.67$ 
with respect to Ref.~\cite{bib:ourPRD}.
This reduction is mostly due to the addition of new data (by $1.30$)
and the inclusion of events with three jets (by $1.25$).
  
The inclusion of \lptj\ events, the addition of newer data, the use of better object identification algorithms, 
and improvements to the analysis technique all decrease the measured asymmetry. 
Both here and in Ref.~\cite{bib:ourPRD}, the analysis strategy was finalized before analyzing Tevatron data.
Together, these changes reduce the measured \afbl\ by $10.5\%$, 
yet no single change accounts for a difference of more than $2.5\%$. 

The $p$-value for the previously published value, 
assuming the asymmetry predicted by \mcatnlo, is $1.7\times10^{-3}$, 
while the $p$-value of the new result is 0.24. These numbers do not account for the systematic
uncertainty on the theoretical predictions.

Most of the asymmetry in the previous analysis is contained in the \lpgefj s 
channel for events with exactly one $b$ tag. 
In the current analysis, the asymmetry in this channel is still high compared to the SM
expectation, 
with $\afbl=\big(16.3 \pm 4.8 \stat  ^{+2.2}_{-1.4} \syst \big)\%$. 
The relative weight of this channel decreased from $\approx50\%$ in the 
previous analysis to $24\%$ in the current analysis.
The \qyl\ distributions in each channel are shown in Fig.~\ref{fig:dy_chan}.
The \afbl\ values of the various channels are compared in Fig.~\ref{fig:chan}, 
Table~\ref{tab:chan_reco} and Table~\ref{tab:chan_prod}. The
consistency between different channels in Fig.~\ref{fig:chan}
corresponds to a $\chi^{2}$ value of 8.1 for three degrees of freedom, 
which corresponds to a tail probability of $4.5\%$. 

\begin{figure}[htbp]
\begin{center}
\includegraphics[width=0.8\linewidth]{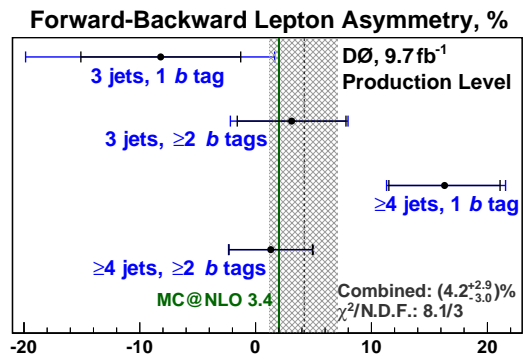}
\end{center}
\vspace{-0.6cm}
\caption{
Measured production\-/level \afbl\ by analysis channel.
The vertical line shows the \mcatnlo\ prediction.
The \chisq\ is of a fit to a single value, shown by
the crosshatched band and the dashed line.
Statistical uncertainties are indicated by the inner vertical lines,
and the total uncertainties by the vertical end lines.
}
\label{fig:chan}
\end{figure}

\begin{table}[htbp]
  \caption{
    Measured and predicted \afbl\ by channel, at reconstruction level.
    \label{tab:chan_reco}
  }
  \begin{ruledtabular}
    \begin{tabular}{lccc}
              &  \multicolumn{2}{c}{\afbl, \%} \\
      Channel & Data & \mcatnlo \\
      \hline \tablestrut 
      \lptj s, 1 $b$ tag          & \hspace{.01em}$-6.8 \pm 6.0 \stat ^{+6.1}_{-5.6} \syst$ & $2.7 \pm 0.4$ \\[0.1cm]
      \lptj s, \getb              & \hspace{.78em}$ 3.7 \pm 4.3 \stat ^{+1.1}_{-1.2} \syst$ & $2.8 \pm 0.3$ \\[0.1cm]
      \lpgefj s, 1 $b$ tag        & \hspace{.28em}$14.8 \pm 4.2 \stat ^{+1.1}_{-1.2} \syst$ & $0.5 \pm 0.3$ \\[0.1cm]
      \lpgefj s, \getb            &               $-0.9 \pm 3.2 \stat ^{+0.3}_{-0.9} \syst$ & $1.1 \pm 0.2$ \\
      \hline
      Total\tablestrut            & \hspace{.79em}$ 2.9 \pm 2.1 \stat ^{+1.5}_{-1.7} \syst$ & $1.6 \pm 0.2$
    \end{tabular}
  \end{ruledtabular}
\end{table}

\begin{table}[htbp]
  \caption{
    Measured \afbl\ by channel at production level.
    The \mcatnlo\ prediction for $\absyl<1.5$ is $\afbl=2.0\%$.
    \label{tab:chan_prod}
  }
  \begin{ruledtabular}
    \begin{tabular}{lc}
      Channel & Measured \afbl, \% \\
      \hline \tablestrut 
      \lptj s, 1 $b$ tag          & \noone        $-8.2 \pm 6.9 \stat ^{+7.0}_{-9.4} \syst$ \\[0.1cm]
      \lptj s, \getb              & \noone\nominus$ 3.1 \pm 4.7 \stat ^{+1.3}_{-2.4} \syst$ \\[0.1cm]
      \lpgefj s, 1 $b$ tag        & \nominus      $16.3 \pm 4.8 \stat ^{+2.2}_{-1.4} \syst$ \\[0.1cm]
      \lpgefj s, \getb            & \noone\nominus$ 1.3 \pm 3.6 \stat ^{+0.8}_{-0.5} \syst$ \\
      \hline
      Total\tablestrut            & \noone\nominus$4.2 \pm 2.3 \stat ^{+1.7}_{-2.0} \syst$  
    \end{tabular}
  \end{ruledtabular}
\end{table}

We also studied the asymmetry in subsamples defined by the charge of the lepton, 
the flavor of the lepton, 
and by the polarities of the \DZ\ magnets, which are reversed every two weeks.
Reversing the magnet polarities greatly reduces possible experimental biases
which involve the lepton.
All measurements agree within at most two standard deviations.

Since the SM-derived corrections to the measured \afbl\ are only 1--2\%, 
the dependence of the results on the SM is small and the results may be validly 
compared to the predictions of beyond the SM predictions.
We tested our analysis method using axigluon samples produced using \madgraph\ combined
with \pythia\ with axigluon masses of $0.2$, $0.4$, $0.8$, and 2\TeV\ with 
completely left-handed, completely right-handed and mixed couplings~\cite{bib:madgraph}. 
In all of these scenarios, the measured production\-/level \afbl\ exhibits additional, 
model-dependent scatter about the simulated \afbl\ values of $<1.5\%\thinspace$(absolute).

%
%
\section{Combination and extrapolation}
\label{sec:comb}
The \DZ\ Collaboration measured \afbl\ in the dilepton channel, with lepton coverage
that extends to $\absyl=2$, finding $\afbl=\big(4.1\pm3.5\stat\pm1.0\syst\big)\%$~\cite{bib:D0dilep}.
To enable a direct combination with the measurements in the \lpj\ channel, the analysis
of Ref.~\cite{bib:D0dilep} is repeated using only leptons with $\absyl<1.5$, finding
$\afbl=\big(4.3\pm3.4\stat\pm1.0\syst\big)\%$. The decrease in the statistical uncertainty is due
to the removal of the events with $\absyl>1.5$, which have a large weight
due to the acceptance corrections and thus increase the statistical uncertainty.

This result is combined with the results of the current measurement using the Best
Linear Unbiased Estimator (BLUE) method, as described in Ref.~\cite{bib:blue}.
Systematic uncertainties are classified by their source as either completely correlated,
\eg, the $b$ tagging uncertainties, or completely uncorrelated, \eg, the background modeling
uncertainties. The combination is a weighted average of the input measurements, with
the dilepton measurement given a weight of $0.43$ and the \lpj\ measurement a weight of $0.57$.
The combined value of \afbl\ for $\absyl<1.5$ is 
$\afbl=\big(4.2\pm2.0\stat\pm1.4\syst\big)\% = \left(4.2\pm2.4\right)\%$.

The measurements are extrapolated to cover the full phase space using the \mcatnlo\ simulation.
Extrapolation adds nothing to our experimental measurements of \afbl, 
but simply extends them by incorporating SM-inspired
predictions for \ttbar\ production outside the lepton rapidity coverage, 
thus facilitating comparison with theoretical calculations and the combination of the measurements.  
Such extrapolated values should not be compared with non-SM predictions~\cite{bib:bsm_acceptance}.
To include dilepton events with $1.5<\absyl<2$ in the extrapolated values,
we extrapolate each result independently before combining them.
If we assume a linear dependence of the asymmetry on \absyl, we find
an extrapolated asymmetry which is proportional to the measured asymmetry,
\begin{equation}
\aext = \afbl \atotmc / \amc,
\end{equation} 
where \atotmc\ is the simulated \afbl\ in the
entire phase space and \amc\ is the simulated \afbl\ within the acceptance.
For the simulated \ttbar\ samples, such an extrapolation overestimates the fully inclusive \afbl.
Furthermore, only the leading order of the \ttbar\ asymmetry has been calculated, and so the
dependence of \afbl\ on \absyl\ is not known precisely.
Therefore we assign a systematic uncertainty to the extrapolation that equals 
the entire effect of the extrapolation.
Using the \mcatnlo\ predictions of $\atotmc/\amc=1.19$ for the \lpj\ measurement
and $1.07$ for the dilepton measurement, and find
a combined extrapolated asymmetry of $\aext = \big(4.7\pm2.3\stat\pm1.5\syst\big)\%$.

%
%
\section{Summary}
Using the full dataset collected by the \DZ\ experiment during Run II of the Tevatron, corresponding to 9.7\ifb,
we have measured the forward-backward asymmetry in the direction of leptons from \ttbar\ events in the \lpj\ channel 
and compared it to a prediction based on \mcatnlo. 
Since the lepton-based asymmetry does not require a full reconstruction of the \ttbar\ event,
this measurement also uses events with only three jets.
The measured asymmetry at production level for $\absyl<1.5$ is $\afbl=\left(4.2^{+2.9}_{-3.0} \right)\%$.

We combined this measurement with the measurement in the dilepton channel and found
a production\-/level asymmetry for $\absyl<1.5$ of $\afbl=\left(4.2\pm2.4\right)\%$,
to be compared to the \mcatnlo\ prediction of $2.0$\%.
We have presented the first measurement of the differential asymmetry as a function of \ptl. 
All results are in agreement with \mcatnlo\ predictions.

%
%
\begin{acknowledgments}
We thank G.~Perez, M.~Mangano, and P.~Skands for enlightening discussions.
%
We thank the staffs at Fermilab and collaborating institutions,
and acknowledge support from the
DOE and NSF (USA);
CEA and CNRS/IN2P3 (France);
MON, NRC KI and RFBR (Russia);
CNPq, FAPERJ, FAPESP and FUNDUNESP (Brazil);
DAE and DST (India);
Colciencias (Colombia);
CONACyT (Mexico);
NRF (Korea);
FOM (The Netherlands);
STFC and the Royal Society (United Kingdom);
MSMT and GACR (Czech Republic);
BMBF and DFG (Germany);
SFI (Ireland);
The Swedish Research Council (Sweden);
and
CAS and CNSF (China).

\end{acknowledgments}

\end{document}